\documentstyle[jair,twoside,11pt,theapa,subfigure,graphicx]{article}

\jairheading{30}{2007}{621-657}{03/07}{12/07} \ShortHeadings{Query-time
Entity Resolution}{Bhattacharya \& Getoor} \firstpageno{621}

\newcommand{\commentout}[1]{%
}
\newcommand{\secref}[1]{Section~\ref{#1}}

\newcommand{\eqnref}[1]{Eq.~(\ref{#1})}
\newcommand{\figref}[1]{Figure~\ref{#1}}
\newcommand{\tabref}[1]{Table~\ref{#1}}

\begin{document}

\title{Query-time Entity Resolution}

\author{\name Indrajit Bhattacharya \email indrajbh@in.ibm.com \\
       \addr IBM India Research Laboratory \\
       Vasant Kunj, New Delhi 110 070, India \\
       \name Lise Getoor \email getoor@cs.umd.edu \\
       \addr Department of Computer Science \\
    University of Maryland,
       College Park, MD 20742 USA}


\maketitle

\begin{abstract}

Entity resolution is the problem of reconciling database references
corresponding to the same real-world entities.
Given the abundance of publicly available
databases that have unresolved entities, we motivate the problem of
\emph{query-time entity resolution}: quick and accurate resolution for
answering queries over such `unclean' databases at query-time.  Since
collective entity resolution approaches --- where related references
are resolved jointly --- have been shown to be more accurate than
independent attribute-based resolution for off-line entity resolution,
we focus on developing new algorithms for collective resolution for
answering entity resolution queries at query-time.  For this purpose,
we first formally show that, for collective resolution, precision and
recall for individual entities follow a geometric progression as
neighbors at increasing distances are considered. Unfolding this
progression leads naturally to a two stage `expand and resolve' query
processing strategy. In this strategy, we first extract the related
records for a query using two novel expansion operators, and then
resolve the extracted records collectively. We then show how the same
strategy can be adapted for query-time entity resolution by
identifying and resolving only those database references that are the
most helpful for processing the query. We validate our approach on two
large real-world publication databases where we show the usefulness of
collective resolution and at the same time demonstrate the need for
adaptive strategies for query processing. We then show how the same
queries can be answered in real-time using our adaptive approach while
preserving the gains of collective resolution. In addition to
experiments on real datasets, we use synthetically generated data to
empirically demonstrate the validity of the performance trends
predicted by our analysis of collective entity resolution over a wide
range of structural characteristics in the data.

\end{abstract}

\section{Introduction}

With the growing abundance of publicly available data in digital form,
there has been intense research on data integration. A critical
component of the data integration process is the entity resolution
problem, where uncertain references in the data to real-world entities
such as people, places, organizations, events, etc., need to be
resolved according to their underlying real-world entities.  Entity
resolution is needed in order to solve the `deduplication'
problem, where the goal is to identify and consolidate pairs of
records or references within the same relational table that are
duplicates of each other. It also comes up as the `fuzzy match'
problem, where tuples from two heterogeneous databases with different
keys, and possibly different schemas, need to be matched and
consolidated.
It goes by different names even within
the data mining and database communities, including record linkage,
object consolidation, and reference reconciliation.

The problem has a long history, and recent years have seen
significant and fruitful research on this problem. However, in spite
of the widespread research interest and the practical nature of the
problem, many publicly accessible databases remain unresolved, or
partially resolved, at best. The popular publication databases,
CiteSeer and PubMed, are representative examples. CiteSeer contains
several records for the same paper or author, while author names in
PubMed are not resolved at all. This is due to a variety of reasons,
ranging from rapid and often uncontrolled growth of the databases
and the computational and other expenses involved in maintaining
resolved entities.

Yet, millions of users access and query such
databases everyday, mostly seeking information that, implicitly or
explicitly, requires knowledge of the resolved entities. For
example, we may query the CiteSeer database of computer science
publications looking for books by `S Russell' \cite{pasula:nips03}.
This query would be easy to answer if all author names in CiteSeer
were correctly mapped to their entities. But, unfortunately, this is
not the case. According to CiteSeer records, Stuart Russell and Peter
Norvig have written more than 100 different books together.
One of the main reasons behind databases containing unresolved entities
is that entity resolution is generally perceived as an expensive
process for large databases. Also, maintaining a `clean' database
requires significant effort to keep pace with incoming records.
Alternatively, we may be searching different online social network
communities for a person named `Jon Doe'. In this case, each online
community may individually have records that are clean. Even then, query
results that return records from all of the sources aggregated together may have
multiple representations for the same 'Jon Doe' entity.
Additionally, in both cases, it is not sufficient to simply return records that match the
query name, `S. Russell' or `Jon Doe' exactly. In order to retrieve
all the references correctly, we may need to retrieve records with
similar names as well, such as 'Stuart Russel' or 'John Doe'.
And, most importantly, for the results to be
useful, we need to partition the records that are returned according
to the real-world entities to which they correspond. Such on-the-fly partitioning of returned results is also necessary when accessing third-party or external databases that do not provide full access possibly due to privacy and other concerns, and can be accessed only via specific query interfaces.

In this paper, we propose an alternative solution for answering entity resolution queries,
where we obviate the need for maintaining resolved entities in a database. Instead, we
investigate entity resolution at query-time, where the goal is to
enable users to query an unresolved or partially resolved database
and resolve the {\em relevant} entities on the fly. A user may
access several databases everyday and she does not want to resolve
all entities in every database that she queries. She only needs to
resolve those entities that are relevant for a particular query. For
instance, when looking for all books by `Stuart Russell' in
CiteSeer, it is not useful to resolve all of the authors in
CiteSeer. Since the resolution needs to be performed at query-time, the requirement is that the resolution process needs to be quick, even if it is not entirely accurate.

Though entity resolution queries have not been addressed in the
literature, there has been significant progress on the general entity
resolution problem. Recent research has focused on the use of
additional relational information between database references to
improve resolution accuracy
\cite{bhattacharya:dmkd04,parag:kdd04-wkshp,dong:sigmod05,ananthakrishna:vldb02,kalashnikov:sdm05}. This improvement is made
possible by resolving related references or records jointly, rather
than independently. Intuitively, this corresponds to the notion that
figuring out that two records refer to the same underlying entity may
in turn give us useful information for resolving other record pairs
that are related. Imagine that we are trying to decide if two authors
`Stuart Russell' and `S Russell' are the same person. We can be more
confident about this decision if we have already decided that their
co-authors `Peter Norvig' and `P. Norvig' are the same person.

As others have done, in our earlier work
\cite{bhattacharya:dmkd04,bhattacharya:tkdd07}, we have
demonstrated using extensive experiments on multiple real and
synthetic datasets that collective resolution significantly improves
entity resolution accuracy over attribute-based and naive relational
baselines. However, its application for query-time entity resolution
is not straight-forward, and this is precisely the problem that we
focus on in this paper. The first difficulty is that collective
resolution works for a database as a whole and not for a specific
query. Secondly, the accuracy improvement comes at a considerable
computation cost arising from the dependencies between related
resolutions. This added computational expense makes its application
in query-time resolution challenging.

In this paper,
which builds on and significantly extends the work presented in
\citeA{bhattacharya:kdd06},
we investigate the application of collective resolution
for queries. First, we formally analyze how accuracies of different
decisions in collective resolution depend on each other and on the
structural characteristics of the data. The recursive nature of the
dependency leads naturally to a recursive `expand and resolve'
strategy for processing queries. The relevant records necessary for
answering the query are extracted by a recursive expansion process and
then collective resolution is performed only on the extracted
records. Using our analysis, we show that the recursive expansion
process can be terminated at reasonably small depths for accurately
answering any query; the returns fall off exponentially as neighbors
that are further away are considered.

However, the problem is that this unconstrained expansion process
can return too many records even at small depths; and thus the query
may still be impossible to resolve in real-time. We address this
issue using an adaptive strategy that only considers the most
informative of the related records for answering any query. This
significantly reduces the number of records that need to be
investigated at query time, but, most importantly, does not
compromise on the resolution accuracy for the query.


Our specific contributions in this paper are as follows:

\begin{enumerate}

\item First, we motivate and formulate the problem of query-time entity
resolution. Our entity resolution approach is based on a relational
clustering algorithm. To the best of our knowledge, clustering based
on queries in the presence of relations has received little attention
in the literature.

\item For collective resolution using relational clustering,
we present an analysis of how the accuracy of different resolution
decisions depends on each other and on the structural
characteristics of the data. We introduce the notion of precision
and recall for individual entities, and show how they follow a
geometric progression as neighbors at increasing distances are
considered and resolved. Our analysis shows that collective use of
relationships can sometimes hurt entity resolution accuracy. This
has not been previously reported in the literature. Our analysis
additionally demonstrates the convergent nature of resolution
performance for the recursive query-resolution strategy that we
propose.

\item For resolving queries collectively, we propose a two-phase
`expand and resolve' algorithm. It first extracts the related
records for a query using two novel expansion operators, and then
resolves the query by only considering the extracted records. We
then improve on this algorithm using an adaptive approach that
selectively considers only the `most informative' ones among the
related records for a query. This enables collective resolution at
query-time without compromising on resolution accuracy for the
query.

\item We present experimental results on two large real-world
datasets where our strategy enables collective resolution in
seconds. We compare against multiple baselines to show that the
accuracy achieved using collective query resolution is significantly
higher than those achieved using traditional approaches.

\item We also use synthetically generated data to demonstrate the gains
of collective query resolution over a wide range of attribute and
relational characteristics. We additionally show that the empirical
results are in agreement with the trends predicted by our analysis
of collective resolution.

\end{enumerate}

The rest of the paper is organized as follows. In
\secref{sec:er-form}, we formalize the relational entity resolution
problem and entity resolution queries, and also illustrate these
with an example. In \secref{sec:rc-er}, we briefly review the
relational clustering algorithm that we employ for collective entity
resolution and then, in \secref{sec:rc-analysis}, investigate how
resolution accuracy for related entities depend on each other for
collective resolution using this algorithm. In
\secref{sec:query-er}, we extend collective resolution for queries,
and describe and analyze an unconstrained recursive strategy for
collectively resolving a query. We then modify this approach in
\secref{sec:adaptive} and present our adaptive algorithm that
extracts only the `most informative' references for resolving a
query. We present experimental results on real and synthetic data in
\secref{sec:exp}, review related work in \secref{sec:rw} and finally
conclude in \secref{sec:concl}.

\section{Entity Resolution and Queries: Formulation}
\label{sec:er-form}

In this section, we formally introduce the entity resolution problem
and also entity resolution queries, and illustrate them using a
realistic example --- that of resolving authors in a citation database
such as CiteSeer or PubMed.

In the simplest formulation of the entity resolution problem, we have
a collection of references, ${\mathcal R}=\{r_i\}$, with attributes
$\{{\mathcal R}.A_{1},\ldots,{\mathcal R}.A_k\}$. Let ${\mathcal E} =
\{e_j\}$ be the unobserved domain entities. For any particular reference $r_i$,
we denote the entity to which it maps as $E(r_i)$. We will say that
two references $r_i$ and $r_j$ are {\em co-referent} if they
correspond to the same entity, $E(r_i)=E(r_j)$. Note that in the case
of an unresolved database, this mapping $E(\mathcal R)$ is \emph{not
provided}. Further, the domain entities ${\mathcal E}$ and even the
number of such entities is not known.  However, in many domains, we
may have additional information about relationships between the
references. To model relationships in a generic way, we use a set of
hyper-edges ${\mathcal H} = \{h_i\}$. Each hyper-edge connects
multiple references. To capture this, we associate a set of references
$h_i.R$ with each hyper-edge $h_i$. Note that each reference may be
associated with zero or more hyper-edges.

Let us now look at a sample domain to see how it can be represented in
our framework. Consider a database of academic publications similar
to DBLP, CiteSeer or PubMed. Each publication in the database has a
set of author names. For every author name, we have a reference
$r_i$ in ${\mathcal R}$. For any reference $r_i$, $r_i.Name$ records
the observed name of the author in the publication. In addition, we
can have attributes such as ${\mathcal R}.Email$ to record other
information for each author reference that may be available in the
paper. Now we come to the relationships for this domain. All the
author references in any publication are connected to each other by
a co-author relationship. This can be represented using a hyper-edge
$h_i\in{\mathcal H}$ for each publication and by having $r_j\in
h_i.R$ for each reference $r_j$ in the publication. If publications
have additional information such as title, keywords, etc, they are
represented as attributes of ${\mathcal H}$.

\begin{figure}[t]
\centering
\includegraphics[angle=0, width=0.8\linewidth]{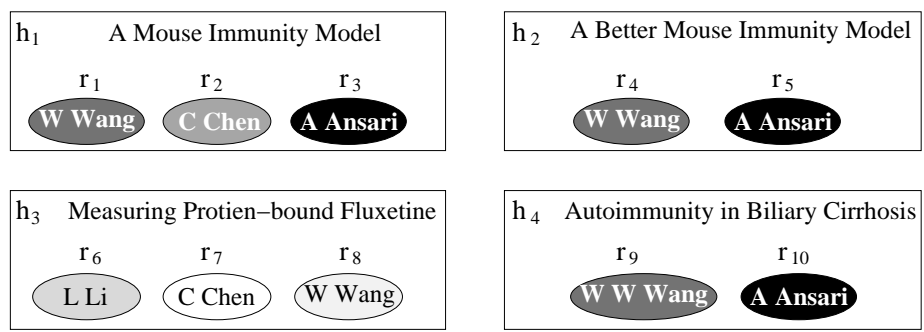}
\caption{An example set of papers represented as references connected
by hyper-edges. References are represented as ovals shaded according
to their entities. Each paper is represented as a hyper-edge (shown as
a rectangle) spanning multiple references.}
\label{fig:example}
\end{figure}
To illustrate, consider the following four papers, which we will use
as a running example:
\begin{enumerate}
\item W. Wang, C. Chen, A. Ansari, ``A mouse immunity model''
\item W. Wang, A. Ansari, ``A better mouse immunity model''
\item L. Li, C. Chen, W. Wang,``Measuring protein-bound fluxetine''
\item W. W. Wang, A. Ansari, ``Autoimmunity in biliary cirrhosis''
\end{enumerate}
To represent them in our notation, we have 10 references $\{
r_1,\ldots, r_{10}\}$ in ${\mathcal R}$, one for each author name,
such that $r_1.Name=$ $\mbox{`W Wang'}$, etc. We also have 4
hyper-edges $\{h_1,\ldots,h_4\}$ in ${\mathcal H}$, one for each
paper. The first hyper-edge $h_1$ connects the three references $r_1$,
$r_2$ and $r_3$ corresponding to the names `W. Wang' , `C. Chen' and
`A. Ansari'. This is represented pictorially in \figref{fig:example}.

Given this representation, the {\bf entity resolution task} is defined
as the partitioning or clustering of the references according to the
underlying entity-reference mapping $E(\mathcal R)$. Two references
$r_i$ and $r_j$ should be assigned to the same cluster if and only if
they are coreferent, i.e., $E(r_i)=E(r_j)$. To illustrate, assume that
we have six underlying entities for our example. This is illustrated
in \figref{fig:example} using a different shading for each entity. For
example, the `Wang's of papers 1, 2 and 4 are names of the same
individual but
the 'Wang' from paper 3 is a reference to a different person.
Also, the
`Chen's from papers 1 and 3 are different individuals. Then, the
correct entity resolution for our example database with $10$
references returns $6$ entity clusters: $\{\{r_1, r_4, r_9\},$
$\{r_8\},\{r_2\},\{r_7\},$ $\{r_3, r_5, r_{10}\},$ $\{r_6\}\}$. The
first two clusters correspond to two different people named `Wang',
the next two to two different people named `Chen', the fifth to
`Ansari' and the last to `Li'.

\commentout{ Correctly resolving all the database references is
important for a number of reasons, such as dealing with redundancy,
correct computation of statistics and discovery of patterns from the
data. But as we have discussed, this is often an expensive process
and as result many databases remain unresolved. However, there are
many applications that need resolved entities, but do not require
that the entire database be entity resolved. Very often, end users
of databases retrieve information using queries. As an example,
consider a CiteSeer user looking up papers written by `S. Russell'.
Since the query involves retrieving references corresponding to a
specific entity, it cannot be answered without the knowledge of
entities. We refer to such queries as {\bf entity resolution
queries}, which we next describe more formally. }

Any query to a database of references is called an {\bf entity
resolution query} if answering it requires knowledge of the
underlying entity mapping $E(\mathcal R)$. We consider two different
types of entity resolution queries. Most commonly, queries are
specified using a particular value $a$ for an attribute ${\mathcal
R}.A$ of the references that serves as a `quasi-identifier' for the
underlying entities. Then the answer to the query $Q(R.A = a)$ should
partition or group all references that have $r.A=a$ according to their
underlying entities. For references to people, the name often serves
as a weak or noisy identifier. For our example bibliographic domain,
we consider queries specified using ${\mathcal R}.Name$. To retrieve
all papers written by some person named `W. Wang', we issue a query
using ${\mathcal R}.Name$ and `W. Wang'. Since names are ambiguous,
treating them as identifiers leads to undesirable results. In this
case, it would be incorrect to return the set $\{r_1, r_4, r_8\}$ of
all references with name `W Wang' as the answer to our query. This
answer does not indicate that $r_8$ is not the same person as the
other two. Additionally, the answer should include the reference
$r_9$ for `W W Wang', that maps to the same entity as the author of
the first paper. Therefore, the correct answer to the entity
resolution query on `W Wang' should be the partition $\{\{r_1, r_4,
r_9\}$$,\{r_8\}\}$.

Entity resolution queries may alternatively be specified using a
specific reference. Imagine a CiteSeer user looking at a paper that
contains some author name.
The user may be interested in looking up other papers written by
the same author, even though they may not know who that author is
precisely.
Then the correct answer to a query on the reference
$r$ is the group of references that are coreferent to $r$, or, in
other words, correspond to the same underlying entity. In our
example, consider a query specified using the reference $r_1$
corresponding to the name `W. Wang' in the first paper. Then the
correct answer to the query is the set of references $\{r_1, r_4,
r_9\}$. To distinguish it from the first type of entity resolution
query, note that it does not include the cluster $\{r_8\}$
corresponding to the other entity that also has name `W. Wang'. This
second query type may be answered by first reducing it to an
instance of the first type as $Q(R.A = r_1.A)$, and then selecting the
entity corresponding to reference $r_1$. We denote this as
$\sigma_{E({\mathcal R})=E(r_1)} (Q(R.A = r_1.A))$. In the rest of this
paper, we focus only on queries of the first type.

\section{Collective Entity Resolution and Relational Clustering}
\label{sec:rc-er}

Although entity resolution for queries has not been studied in the
literature, the general entity resolution problem has received a lot
of attention. We review related work in detail in \secref{sec:rw}.
In this section, we briefly review the different categories of
proposed approaches before discussing how they may be adapted for
query-time entity resolution.

In most entity resolution applications, data labeled with the
underlying entities is hard to acquire. Our focus is on unsupervised
approaches for resolving entities. Traditionally, attributes of
individual references, such as names, affiliation, etc., for person
references, are used for comparing references. A similarity measure
is generally employed over attributes, and only those pairs of
references that have attribute similarity above a certain threshold
are considered to be co-referent. This {\bf attribute-based entity
resolution} approach ({\bf A}) often runs into problems. In our
example, it is hard to infer with just attributes that references
$r_1$ and $r_8$ are not co-referent although they have the same
name, while $r_1$ and $r_9$ {\em are} co-referent although their
names are different.

When relations between references are available, they may also be
taken into account for computing similarities in the {\bf naive
relational entity resolution} approach ({\bf NR})
\shortcite{ananthakrishna:vldb02,bhattacharya:tkdd07}. For computing
similarities between two references, this approach additionally
considers the attributes of the related references when comparing
the attributes of their related references. In our example, this
approach returns a higher similarity between $r_1$ (`W. Wang') and
$r_9$ (`W. W. Wang') than the attribute-based approach, since they
have co-authors $r_3$ and $r_{10}$ with very similar (identical, in
this case) names.
Although
this approach can improve performance in
some cases, it does not always work. For instance, the two `W. Wang'
references $r_1$ and $r_8$ are not co-referent, though they both
have co-authors with identical names `C. Chen'.

Instead of considering the attribute similarities of the related
references, the {\bf collective entity resolution} approach
\cite{pasula:nips03,bhattacharya:dmkd04,parag:kdd04-wkshp,mccallum:nips04,li:aimag05,dong:sigmod05,kalashnikov:sdm05}
takes into account the {\em resolution decisions} for them. In our
previous example, the correct evidence to use for the pair of
references $r_1$ and $r_8$ is that their co-author references do not
map to the same entity, although they have similar names.
Therefore, in order to resolve the `W. Wang' references in the
collective resolution approach, it is necessary to {\em resolve} the
`C. Chen' references as well, instead of considering the similarity
of their attributes. The collective entity resolution approach has
recently been shown to improve entity resolution accuracy over the
previous approaches but is computationally more challenging. The
references cannot be resolved independently. Instead, any resolution
decision is affected by other resolutions through hyper-edges.

In earlier work
\shortcite{bhattacharya:dmkd04,bhattacharya:wiley06,bhattacharya:tkdd07},
we developed a relational clustering algorithm ({\bf RC-ER})
for collective entity resolution using relationships.
The goal of this approach is to cluster
the references according to their entities taking the relationships
into account. We associate a cluster label $r.C$ with each reference
to denote its current cluster membership. Starting from an initial
set of clusters ${\mathcal C}=\{c_i\}$ of references, the algorithm
iteratively merges the pair of clusters that are the most similar.
To capture the collective nature of the cluster assignment, the
similarity measure between pairs of clusters considers the cluster
labels of the related references. The similarity of two clusters
$c_i$ and $c_j$ is defined as a linear combination of their attribute
similarity $sim_A$ and their relational similarity $sim_R$:
\begin{equation}
sim(c_i,c_j) = (1-\alpha)\times sim_{A}(c_i,c_j) +
\alpha \times sim_{R}(c_i,c_j)
\label{eqn:alpha}
\end{equation}
where $\alpha$ ($0\leq\alpha\leq 1$) is the combination weight. The
interesting aspect of the collective approach is the dynamic nature of
the relational similarity. The similarity between two references
depends on the {\em current} cluster labels of their related
references, and therefore changes when related references change
clusters. In our example, the similarity of the two clusters
containing references `W. Wang' and `W. W. Wang' increases once their
co-author references named `A. Ansari' are assigned to the same
cluster. We now briefly review how the two components of the
similarity measure are defined.

\paragraph{Attribute Similarity:} For each reference attribute,
we use a similarity measure that returns a value between $0$ and $1$
for two attribute values indicating the degree of similarity between
them. Several sophisticated similarity measures have been developed
for names, and popular TF-IDF schemes may be used for other textual
attributes such as keywords. The measure that works best for each
attribute may be chosen. Finally, a weighted linear combination of the
similarities over the different attributes yields the combined
attribute similarity between two reference clusters.

\paragraph{Relational Similarity:} Relational similarity between two
clusters considers the similarity of their `cluster neighborhoods'.
The neighborhood of
each cluster is defined by the hyper-edges associated with the
references in that cluster. Recall that each reference $r$ is
associated with one or more hyper-edges in ${\mathcal H}$.
Therefore, the hyper-edge set $c.H$ for a cluster $c$ of references
is defined as
\begin{equation}
c.H = \bigcup_{r \in {\mathcal R} \wedge r.C = c} \{h \mid
h\in{\mathcal H}\ \wedge r\in h.R \}
\end{equation}
This set defines the hyper-edges that connect a cluster $c$ to other
clusters, and are the ones that relational similarity needs to
consider. To illustrate, when all the references in our running
example have been correctly clustered as in \figref{fig:example}(b),
the hyper-edge set for the larger `Wang' cluster is $\{h_1, h_2,
h_4\}$, which are the hyper-edges associated with the references
$r_1$, $r_4$ and $r_9$ in that cluster.

Given the hyper-edge set for any cluster $c$, the neighborhood
$Nbr(c)$ of that cluster $c$ is the set of clusters labels of the
references spanned by these hyper-edges:
\begin{equation}
Nbr(c) = \bigcup_{h\in c.H, r\in h} \{ c_j \mid c_j=r.C \}
\end{equation}
For our example `Wang' cluster, its neighborhood consists of the
`Ansari' cluster and one of the `Chen' clusters, which are connected
by its edge-set. Then, the relational similarity measure between two
clusters, considers the similarity of their cluster neighborhoods.
The neighborhoods are essentially sets (or multi-sets) of cluster
labels and there are many possible ways to define the similarity of
two neighborhoods \shortcite{bhattacharya:tkdd07}. The specific similarity measure that we use for our experiments in this paper is Jaccard similarity\footnote{Jaccard
similarity for two sets $A$ and $B$ is defined as $ Jaccard(A,B) =
\frac{ |A\cap B| } { |A\cup B| }$} :
\begin{equation}
sim_{R}(c_i, c_j) = Jaccard(Nbr(c_i),Nbr(c_j))
\end{equation}

\paragraph{Clustering Algorithm:}
Given the similarity measure for a pair of clusters, a greedy
relational clustering algorithm
can be used for collective
entity resolution. \figref{fig:rc-algo} shows high-level pseudo-code
for the complete algorithm.
\begin{figure}[t]
\begin{minipage}[c]{5in}
\begin{tabbing}
aaaaaaaaaaaa \= aaaaaa \= aaaaaaaa \= aaaaaa \=> \kill
   \>  Algorithm RC-ER (Reference set $\mathcal R$)\\
1. \>  Find similar references in $\mathcal R$ using blocking \\
2. \>  Initialize clusters using bootstrapping \\
\\
3. \> For clusters $c_i,c_j$ such that similar$(c_i,c_j)$ \\
4. \> \>    Insert $\langle sim(c_i,c_j),c_j,c_j \rangle$ into priority queue \\\\
5. \> While priority queue not empty \\
6. \> \>    Extract $\langle sim(c_i,c_j),c_i,c_j\rangle$ from queue \\
7. \> \>    If $sim(c_i,c_j)$ less than threshold, then stop \\
8. \> \>    Merge $c_i$ and $c_j$ to new cluster $c_{ij}$ \\
9. \> \>    Remove entries for $c_i$ and $c_j$ from queue \\
10.\> \>    For each cluster $c_k$ such that similar$(c_{ij},c_k)$ \\
11.\> \> \>     Insert $\langle sim(c_{ij},c_k),c_{ij},c_k\rangle$ into queue \\12.\> \>    For each cluster $c_n$ neighbor of $c_{ij}$ \\
13.\> \> \>      For $c_k$ such that similar$(c_k,c_n)$ \\
14.\> \> \> \>         Update $sim(c_k,c_n)$ in queue \\
\end{tabbing}
\normalsize
\end{minipage}
\caption{High-level description of the relational clustering algorithm}
\label{fig:rc-algo}
\end{figure}
The algorithm first identifies the candidate set of potential
duplicates using a `blocking' approach
\cite{hernandez:sigmod95,monge:dmkd97,mccallum:kdd00}. Next, it initializes the
clusters of references, identifies the `similar' clusters --- or
potential merge-candidates --- for each cluster, inserts all the
merge-candidates into a priority queue and then iterates over the
following steps. At each step, it identifies the current `closest
pair' of clusters from the candidate set and merges them to create a
new cluster. It identifies new candidate pairs and updates the
similarity measures for the `related' cluster pairs. This is the key
step where evidence flows from one resolution decision to
other related ones and this distinguishes relational clustering from
traditional clustering approaches. The algorithm terminates when the
similarity for the closest pair falls below a threshold or when the
list of potential candidates is exhausted. The algorithm is
efficiently implemented to run in $O(nk\log n)$ time for $n$
references where each `block' of similar names is connected to $k$
other blocks through the hyper-edges.

\subsection{Issues with Collective Resolution for Queries}
In previous work, we (and others) have shown that
collective resolution using relationships improves entity resolution
accuracy significantly for offline cleaning of databases. So,
naturally, we would like to use the same approach for query-time
entity resolution as well. However, while the attribute-based and
naive relational approaches discussed earlier can be applied at
query-time in a straight-forward fashion, that is not the case for
collective resolution. Two
issues come up when using
collective resolution for queries. First, the set of references that
influence the resolution decisions for a query need to be
identified. When answering a resolution query for `S. Russell' using
the attribute-based approach, it is sufficient to consider all
papers that have `S. Russell' (or, similar names) as author name.
For collective resolution, in contrast, the co-authors of these
author names, such as `P. Norvig' and `Peter Norvig', also need to be
clustered according to
their
entities. This in turn requires clustering
their co-authors and so on. So the first task is to analyze these
dependencies for collective resolution and identify the references
in the database that are relevant for answering a query. But this is
not enough. The set of references influencing a query may be
extremely large, but the query still needs to be answered quickly
even though the answer may not be completely accurate. So the second
issue is performing the resolution task at query-time. These are the
two problems that we address in the next few sections.

\section{Analysis of Collective Resolution using Relational Clustering}
\label{sec:rc-analysis}

For collective entity resolution, we have seen that resolution
performance for the query becomes dependent on the resolution
accuracy of the related entities. Before we can analyze which other
references influence entity resolution for the query and to what
extent, we need to analyze the nature of this dependence for
collective resolution in general. In this section, we identify the
structural properties of the data that affect collective entity
resolution and formally model the interdependent nature of the
resolution performance. This analysis also helps us to understand
when collective resolution using relational clustering helps, and,
equally importantly, when it has an adverse effect as compared
against traditional attribute-based resolution.

The goal of an entity resolution algorithm is to partition the set
${\mathcal R} = \{r_i\}$ of references into a set of clusters
${\mathcal C} = \{c_i\}$ according to the underlying entities
${\mathcal E} = \{e_i\}$. The accuracy of the resolution depends on
how closely the separation of the references into clusters
corresponds to the underlying entities. We consider two different
measures of performance. The first measure is {\em recall} for each
entity. For any entity $e_i$, recall counts how many pairs of
references corresponding to $e_i$ are correctly assigned to the same
computed cluster. The second measure is {\em precision} for each
computed cluster. For any cluster $c_i$, precision counts how many
pairs of references assigned to $c_i$ truly correspond to the same
underlying entity. (Alternatively, {\em imprecision} measures how many pairs of references assigned to the cluster do not correspond to the same entity.) In the next two subsections, we analyze how these
two performance metrics are influenced, first, by the attribute
values of the references, and then, by the observed relationships
between them.

\subsection{Influence of Attributes}
First, consider an entity resolution algorithm that follows the
traditional attribute-based approach and
the analysis of its performance.
Such an algorithm only considers the attributes of individual
references. It uses a similarity measure defined over the domain of
attributes, and considers pair-wise attribute similarity between
references for resolving them. Let us define two references to be
$\epsilon$-similar if their attribute-similarity is at least $\epsilon$.
Then, given a resolution threshold $\epsilon$, the attribute-based
approach assigns a pair of references to the same cluster if and
only if they are $\epsilon$-similar. To illustrate using our example,
using any similarity measure defined over names and an appropriately
determined similarity threshold $\epsilon$, the attribute-based
approach would assign the three `W. Wang' references ($r_1$, $r_4$,
$r_8$) to one cluster $c_1$ and the `W. W. Wang' reference ($r_9$)
to a different cluster $c_2$. This resolution of the Wang references
is not perfect in terms of precision or recall, since references
$r_1$, $r_4$ and $r_9$ map to one entity $e_1$ and $r_8$ maps to a
second entity $e_2$. Cluster $c_1$ has precision less than 1, since
it incorrectly includes references for two different entities, and
recall is less than 1 for entity $e_1$, since its references are
dispersed over two different clusters.

In order to analyze the performance of this attribute-based
resolution approach given an arbitrary dataset, we now characterize
a dataset in terms of the attribute values of its references.
Intuitively, the attribute-based approach works well when the
references corresponding to the same entity are similar in terms of
their attributes, and when the references corresponding to different
entities are not. To capture this formally, we define two
probabilities that measure the attribute-similarity of references
that map to the same entity, and the attribute-similarity of those
that map to different entities:
\begin{itemize}
\item {\bf attribute identification
probability} ${\bf a_I}(e, \epsilon)$: the probability that a pair of
references chosen randomly from those corresponding to entity $e$
are $\epsilon$-similar to each other.
\item {\bf attribute ambiguity probability} ${\bf a_A}(e_1, e_2, \epsilon)$:
the probability that a pair of references chosen randomly such that
one corresponds to entity $e_1$ and the other to entity $e_2$ are
$\epsilon$-similar to each other.
\end{itemize}

To illustrate using the four `Wang' references, $r_1$, $r_4$ and
$r_9$ correspond to the same entity $e_1$ and $r_8$ corresponds to a
different entity $e_2$. Also, assume that for some similarity
measure for names and an appropriate threshold $\epsilon$, references
$r_1$, $r_4$ and $r_8$ are $\epsilon$-similar to each other. Then, of
the 3 pairs of references corresponding to entity $e_1$, only one
($r_1$ and $r_4$) is $\epsilon$-similar, so that the attribute
identification probability $a_I(e_1,\epsilon)$ for entity $e_1$ is
0.33. On the other hand, of the three pairs of references such that
one maps to $e_1$ and the other to $e_2$, two ($r_1$ and $r_8$,
$r_4$ and $r_8$) are $\epsilon$-similar. This means that the attribute
ambiguity probability $a_A(e_1,e_2,\epsilon)$ between $e_1$ and $e_2$
is 0.66.

As can be seen from the above example, the performance of the
attribute-based clustering algorithm can be represented in terms of
these two probabilities. For any specified threshold $\epsilon$, the
pairs of references for any entity that are correctly recalled are
the ones that are $\epsilon$-similar, which is exactly what $a_I(e,
\epsilon)$ captures. Therefore, the recall for any domain entity $e$
is $R(e, \epsilon)=a_I(e, \epsilon)$. On the other hand, consider the
cluster assignment for all the references that correspond to two
entities $e_1$ and $e_2$. The pairs that are incorrectly clustered
together are those that correspond to two different entities, and
yet are $\epsilon$-similar. This is what $a_A(e_1,e_2,\epsilon)$
captures. Therefore the imprecision of the cluster assignment of
reference pairs corresponding to entities $e_1$ and $e_2$ is
$I(e_1,e_2,\epsilon)=a_A(e_1,e_2,\epsilon)$. Alternatively, the
precision is given by $P(e_1,e_2,\epsilon)\equiv
1-I(e_1,e_2,\epsilon)=1-a_A(e_1,e_2,\epsilon)$.

\subsection{Influence of Relationships}

Now, we consider the collective entity resolution approach that
additionally makes use of relationships, and analyze its impact on
entity resolution accuracy. Recall that we have a set ${\mathcal
H}=\{h_j\}$ of observed co-occurrence relationships between the
references. Such co-occurrences between references are useful for
entity resolution when they result from strong ties or relations
between their underlying entities. Specifically, we assume that
references to any entity $e_i$ co-occur frequently with references
to a small set of other entities $\{e_i^1,\ldots, e_i^k\}$, which we
call the entity neighbors, denoted $N(e_i)$, of entity $e_i$.

Assuming such a neighborhood relationship among the underlying
entities allows us to analyze the performance of the relational
clustering approach. For those reference
pairs that are $\epsilon$-similar in terms of attributes, the attribute
evidence is enough for resolution. But now, unlike attribute-based clustering, any pair of references that are $\delta$-similar in terms of attributes, for some
$\delta<\epsilon$, are considered as candidates for being
clustered together. Not all of them actually get assigned to the same cluster. For
reference pairs that are in the ring of uncertainty between $\epsilon$
and $\delta$, their relationships play a role in determining if they
are similar enough, and consequently, if they should be clustered together. Specifically, if references $r_i$ and $r_j$
co-occur through hyper-edge $h$ and references $r'_i$ and $r'_j$
co-occur through hyper-edge $h'$, then the relational similarity of
the pair ($r_i$, $r'_i$) is more when ($r_j$, $r'_j$) belong
to the same cluster. In general, multiple such relationships may be
needed for tipping the balance, but for simplicity, we assume for
now that a single pair of related references is sufficient. In other
words, $r_i$ and $r'_i$ get assigned to the same cluster if $r_j$
and $r'_j$ are in the same cluster.

\begin{figure}[t]
\centering
\subfigure{
\includegraphics[angle=0, width=0.45\linewidth]{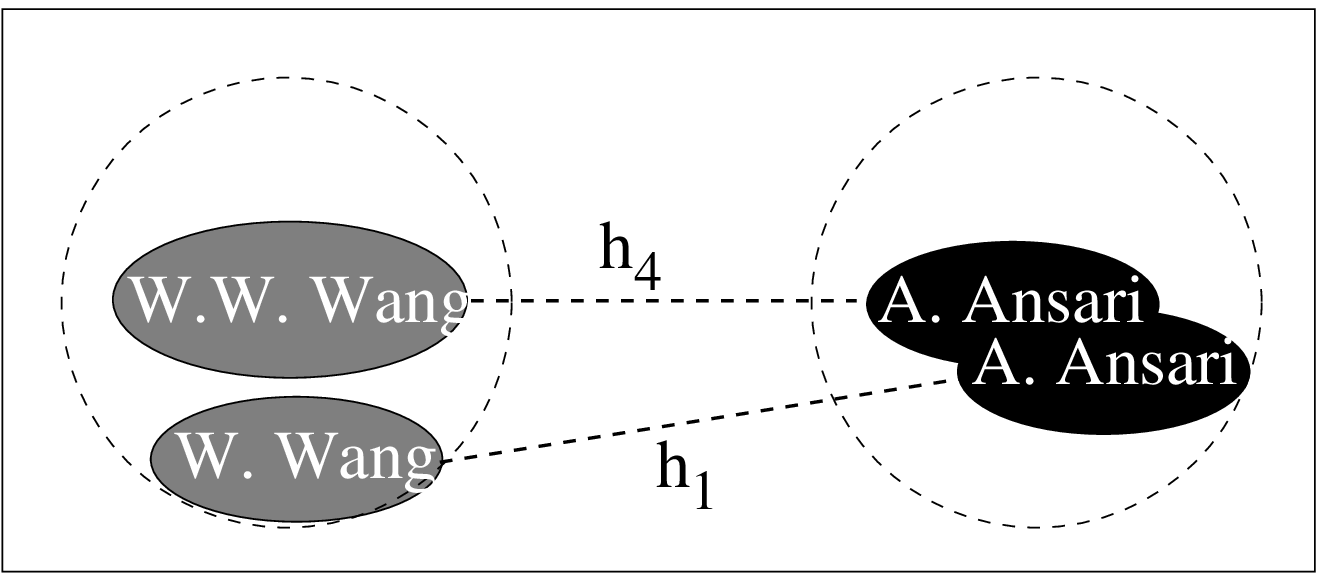}
}
\subfigure{
\includegraphics[angle=0, width=0.45\linewidth]{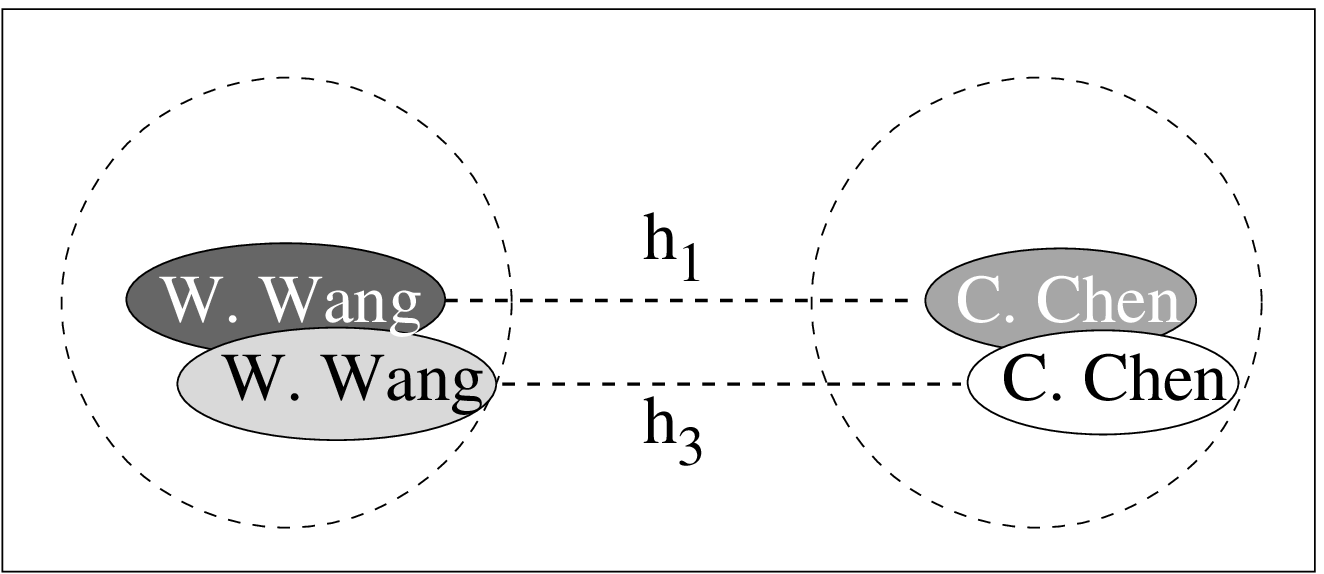}
}
\caption{Illustration of (a) identifying relation and (b) ambiguous
relation from running example. Dashed lines represent co-occurrence
relations.}
\label{fig:rel}
\end{figure}

We now analyze the impact that this approach has on entity
resolution performance. Without loss of generality, assume that the
$(r_j,r'_j)$ pair get clustered together first by the relational
clustering algorithm. This results in the other pair $(r_i,r'_i)$
also getting clustered at some later iteration by considering this
relational evidence. To see if this is accurate, we consider two
situations, as we did with attribute evidence. The first is shown in
\figref{fig:rel}(a), where both pairs truly correspond to the same
entity. Then the collective resolution decision is correct and we
say that hyper-edges $h$ and $h'$ are {\em identifying
relationships} for that entity. Formally,
\begin{eqnarray}
IRel(h,h',e) & \equiv & \exists\ r_i,r_j\in h.R,\ r'_i,r'_j\in h'.R, \nonumber \\
& & E(r_i)=E(r'_i)=e,\ E(r_j)=E(r'_j)
\end{eqnarray}
On the other hand, we may have a different scenario, in which both
pairs of references correspond to two different entities. This
second scenario is depicted in \figref{fig:rel}(b). Then the first
decision to resolve $(r_j,r'_j)$ as co-referent is incorrect, and
relational evidence obtained through hyper-edges $h$ and $h'$
consequently leads to the incorrect resolution of $(r_i,r'_i)$. In
this situation, collective resolution hurts accuracy, and we say
that $h$ and $h'$ form {\em ambiguous relationships} for both pairs
of entities, whose references may be incorrectly clustered as a
result of these relationships. Formally,
\begin{eqnarray}
IAmb(h,h',e,e') & \equiv & \exists\ r_i,r_j\in h.R,\ r'_i,r'_j\in h'.R, \nonumber \\
& & E(r_i)=e,\ E(r'_i)=e',\ e\neq e', \nonumber \\
& & E(r_j)\neq E(r'_j)
\end{eqnarray}

In general, a reference $r_i$ can have a co-occurrence relation $h$
that includes more than one other reference. We may think of this as
multiple co-occurrence pairs involving $r_i$. Cluster labels of all
these other references in the pairs influence resolution decisions
for $r_i$. When resolving $r_i$ with another reference $r'_i$ that
participates in co-occurrence relation $h'$, the fraction of common
cluster labels between $h$ and $h'$ determines whether or not $r_i$
and $r'_i$ will be clustered together. If they are assigned to the
same cluster, $h$ and $h'$ are labeled identifying or ambiguous
relationships based on whether $r_i$ and $r'_i$ are actually
co-referent or not.

Formally, we define:
\begin{itemize}
\item {\bf identifying relationship probability} ${\bf r_I}(e, \delta)$:
the probability that a randomly chosen pair of $\delta$-similar references
corresponding to entity $e$ has identifying relationships $h$ and
$h'$ with some other entity.
\item {\bf ambiguous relationship probability} ${\bf r_A}(e_1,e_2,\delta)$:
the probability that a pair of $\delta$-similar references, chosen randomly such that one corresponds to entity $e_1$ and the other to entity $e_2$, has
ambiguous relationships $h$ and $h'$ with some other pair of
entities.
\end{itemize}

To illustrate these probabilities using our example, we have two
`Wang' entities, $e_1$ that has references $r_1$, $r_4$ and $r_9$,
and $e_2$ that has reference $r_8$. Assume that the attribute
threshold $\delta$ is such that all six pairs are considered
potential matches. Of the three pairs of references corresponding to
$e_1$, all of them have identifying relationships with the `Ansari'
entity. So, $r_I(e_1,\delta)=1$. To measure the relational ambiguity
between the two `Wang' entities, we consider the 3 possible pairs
($r_1$ and $r_8$, $r_4$ and $r_8$, $r_9$ and $r_8$). Of these only
one ($r_1$ and $r_8$) pair has ambiguous relationships with two
different `Chen' entities. So, $r_A(e_1,e_2,\delta)=0.33$.

Given these two probabilities, we can analyze the performance of our
relational clustering algorithm that combines attribute and
relational evidence for collective entity resolution. It is not hard
to see that the recall for any entity depends recursively on the
recall of its neighbor entities. Any pair of references for entity
$e$ is resolved correctly on the basis of attributes alone with
probability $a_I(e,\epsilon)$ (the identifying attribute
probability). Furthermore, it may still be resolved correctly in the
presence of identifying relationships with a neighbor entity, {\em
if the related reference pair for the neighbor is resolved
correctly}. Denoting as $R(e,\epsilon,\delta)$ the recall for entity
$e$ and that for its neighbors as $R(N(e),\epsilon,\delta)$, we
have:
\begin{equation}
R(e,\epsilon,\delta) = a_I(e,\epsilon) + (1-a_I(e,\epsilon))\times
r_I(e,\delta)\times R(N(e),\epsilon,\delta)
\end{equation}

On the other hand, consider a pair of entities $e_1$ and $e_2$. The
cluster assignment for a pair of references corresponding to $e_1$
and $e_2$ is imprecise on the basis of its attributes alone with
probability $a_A(e_1,e_2,\epsilon)$. Even otherwise, the cluster
assignment can go wrong by considering relational evidence. This
happens in the presence of ambiguous relationships with references
corresponding to another pair of entities, {\em if those references
are also clustered together incorrectly}. So the imprecision
$I(e_1,e_2,\epsilon,\delta)$ of the cluster assignment of reference
pairs corresponding to entities $e_1$ and $e_2$ turns out to be:
\begin{equation}
I(e_1,e_2,\epsilon,\delta) = a_A(e_1,e_2,\epsilon) +
(1-a_A(e_1,e_2,\epsilon))\times r_A(e_1,e_2,\delta)\times
I(N(e_1),N(e_2),\epsilon,\delta)
\end{equation}

In general, any entity $e$ has multiple neighbors $e^i$ in its neighborhood $N(e)$. To
formalize the performance dependence on multiple neighbors, assume
that if a co-occurrence involving references corresponding to $e$ is
chosen at random, the probability of selecting a co-occurrence with
a reference corresponding to $e^i$ is $p^e_i$. Then recall is given
as:
\begin{equation}
R(e) = a_I(e) + (1-a_I(e))\times r_I(e)\times \sum_{i=1}^{|N(e)|}p^e_i
R(e^i) \label{eqn:rec}
\end{equation}
Note that we have dropped $\epsilon$ and $\delta$ for notational
brevity. For defining imprecision, observe that a reference
corresponding to any neighbor $e_1^i$ of $e_1$ may co-occur with a
reference for any neighbor $e_2^j$ of $e_2$ with probability
$p^{e_1}_i p^{e_2}_j$. Then imprecision is given as:
\begin{equation}
I(e_1,e_2) = a_A(e_1,e_2) + (1-a_A(e_1,e_2))\times
r_A(e_1,e_2)\times \sum_{i=1}^{|N(e_1)|}\sum_{j=1}^{|N(e_2)|}p^{e_1}_i p^{e_2}_j
I(e_1^i,e_2^j) \label{eqn:imp}
\end{equation}

\commentout{
The above equations bring out the dependence structure in entity
resolution performance resulting from collective resolution. The
accuracy for any entity depends directly on the accuracies for all
its neighbors. The parameters of the dependency are properties of
both the attributes and relationships of the underlying entities.
Additionally, the equations enable us to analyze the advantages and
disadvantages of relational clustering for entity resolution.
}

Given similarity thresholds $\epsilon$ and $\delta$,
relational clustering increases recall
beyond that achievable using attributes alone. This improvement is
larger when the probability of identifying relationships is higher.
On the flip side, imprecision also increases with relational
clustering.
 Typically, a low attribute
threshold $\epsilon$ that corresponds to high precision is used, and
then recall is increased using relational evidence. When the
probability of ambiguous relations $r_A$ is small, the accompanying
increase in imprecision is negligible, and performance is improved
overall. However, the higher the ambiguous relationship probability
$r_A$, the less effective is relational clustering. Thus the balance
between ambiguous and identifying relations determines the overall
benefit of collective resolution using relational clustering. When
$r_A$ is high compared to $r_I$, imprecision increases faster than
recall, and overall performance is adversely affected compared to
attribute-based clustering. \eqnref{eqn:rec} and \eqnref{eqn:imp} quantify this dependence of resolution performance for any entity on the nature of its relationships with other entities. In the next section, we will use these equations to design and analyze a relational clustering algorithm for answering entity resolution queries.

\section{Collective Resolution for Queries} \label{sec:query-er}

Our analysis of collective resolution using relational clustering
showed that the resolution accuracy for any underlying entity
depends on the resolution accuracy for its related/neighboring
entities. For the problem of answering entity resolution queries,
the goal is not to resolve all the entities in the database. We need
to resolve entities for only those references that are retrieved for
the query. We have seen that collective resolution leads to
potential performance improvements over attribute-based resolution.
We now investigate how collective resolution can be applied for
answering queries to get similar improvements. The obvious hurdle is
illustrated by the expressions for performance metrics in
\eqnref{eqn:rec} and \eqnref{eqn:imp}. They show that in order to
get performance benefits for resolving the query using relational
clustering, we need to resolve the neighboring entities as well.
Furthermore, to resolve the neighboring entities, we need to resolve
their neighboring entities, and so on. These other entities that
need to be resolved can be very large in number, and resolving them is
expensive in terms of query-processing time.  Also, none of them are
actually going to be retrieved as part of the answer to the query.
So it is critical to identify and resolve those entities that
contribute the most for improving resolution accuracy for the query.
We propose a two-stage query processing strategy, consisting of an
{\em extraction phase}, for identifying all the relevant references
that need to be resolved for answering the query, and a {\em
resolution phase}, where the relevant references that have been
extracted are collectively resolved using relational clustering.
Unfolding \eqnref{eqn:rec} and \eqnref{eqn:imp} starting from the
query entities leads to a natural expansion process. In this
section, we describe the extraction process using two novel
expansion operators and, in parallel, we analyze the improvement in
resolution accuracy that is obtained from considering
co-occurrences.

Recall that an entity resolution query $Q(R.A=a)$ is specified using
an attribute $A$ and a value $a$ for it. The answer to the query
consists of a partitioning of all references $r$ that have $r.A=a$
or some value $\delta$-similar to $a$. The correct answer to the
query, in general, involves references from multiple entities
$\{e_q\}$. We measure resolution accuracy for the query using two
metrics as before. For each of the query entities $e_q$, we measure
recall $R(e_q)$ and imprecision $I(e_q,e')$ with respect to any
other entity $e'$. Entity $e'$ may or may not belong to $\{e_q\}$.

Before going into the details of our algorithm for collective
resolution of queries, we briefly recall the accuracy of the
attribute-based strategy of resolving a query. This approach
considers all references $r$ with $r.A$ $\delta$-similar to $a$, and
resolves them using their attributes only. The recall that results
from this approach is $R(e_q,\delta)=a_I(e_q,\delta)$, and the
imprecision is given by $I(e_q,e',\delta)=a_A(e_q,e',\delta)$.

\begin{figure}[t]
\centering
\includegraphics[angle=0, width=0.7\linewidth]{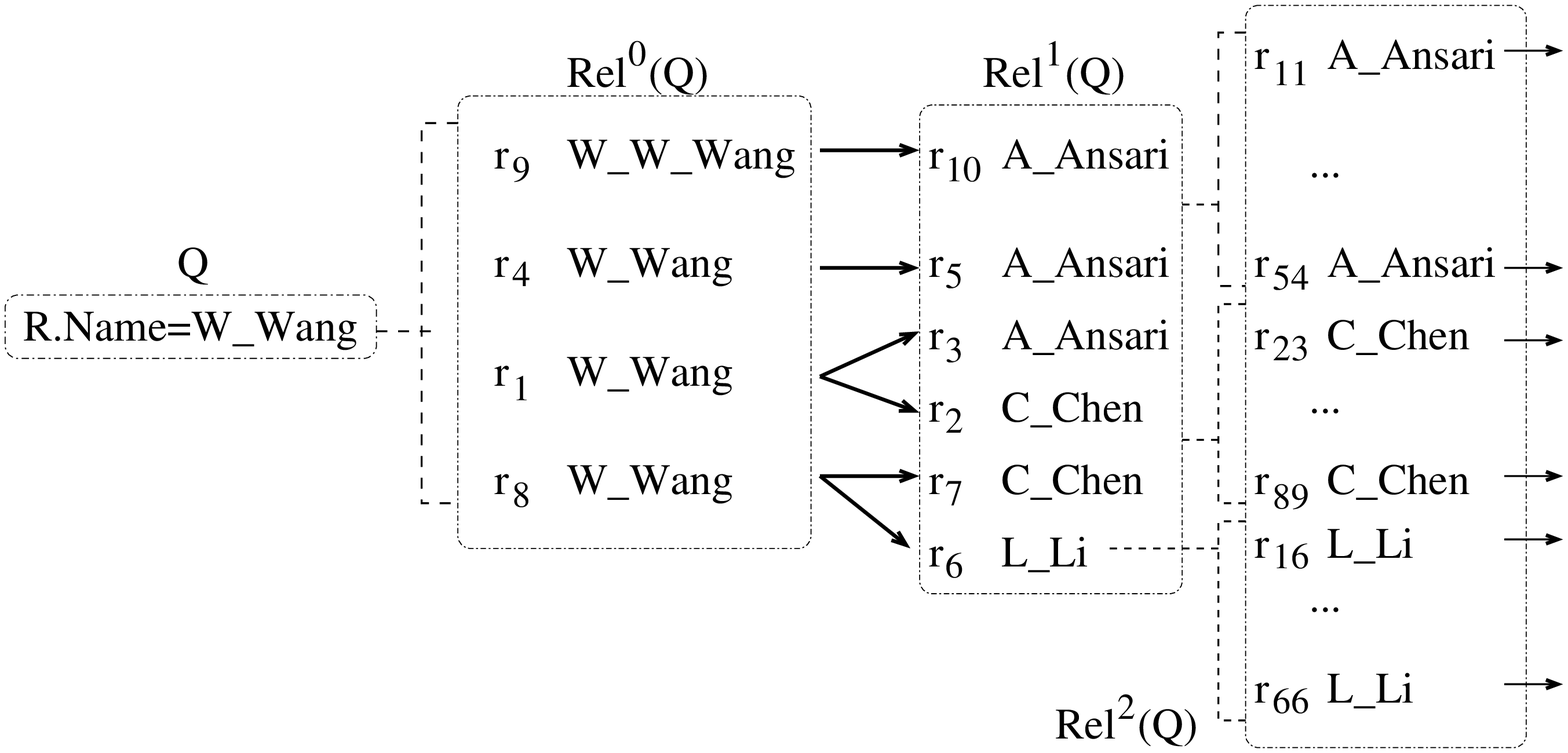}
\caption{Relevant set for query $Q({\mathcal R}.Name = W. Wang)$
using H-expansion and A-expansion alternately} \label{fig:expansion}
\end{figure}

We propose two expansion operators for constructing the relevant set
for an entity resolution query. We denote as {\em level-0
references} all references that are $\delta$-similar to the query
attribute. These are the references that the user is interested in,
and the goal is to resolve these correctly. The first operator we
introduce is the {\bf attribute expansion operator} $X_A$, or
A-expansion for short. Given an attribute $A$ and a value $a$ for
that attribute, $X_A(a,\delta)$ returns all references $r$ whose
attributes $r.A$ exactly match $a$ or are $\delta$-similar to $a$.
For a query $Q(R.A=a)$, the {\em level-0} references can be retrieved
by expanding $Q$ as:
\begin{displaymath}
Rel^0(Q) = X_A(a,\delta)
\end{displaymath}
The first step in \figref{fig:expansion} shows A-expansion for the
query $Q({\mathcal R}.Name = W. Wang)$ in our example. It retrieves
the four references ($r_1$,$r_4$,$r_8$,$r_9$) that have name `W.
Wang' or `W. W. Wang'.

To consider co-occurrence relations, we construct the {\em level-1}
references by including all references that co-occur with {\em
level-0} references. For this, we use our second operator, which we
call {\bf hyper-edge expansion} $X_H$, or H-expansion. For any
reference $r$, $X_H(r)$ returns all references that share a
hyper-edge with $r$, and for a set $R$ of references $X_H(R)$
returns $\bigcup_{r\in R}X_H(r)$. Collective entity resolution
requires that we consider all co-occurring references for each
reference. This is achieved by performing H-expansion on the
references at {\em level-0} to retrieve the {\em level-1}
references:
\begin{displaymath}
Rel^1(Q)=X_H(Rel^0(Q))
\end{displaymath}
\figref{fig:expansion} illustrates this operation in our example,
where $X_H(r_1)$ retrieves references `C.  Chen' ($r_2$) and
`A. Ansari' ($r_3$), and so on.

To perform collective resolution for the query, we additionally need
to {\em resolve} the references at {\em level-1}. One option for {\em
level-1} references is attribute-based resolution using a conservative
$\epsilon$-similarity to keep imprecision to a minimum.  We can use
our analysis technique from before to evaluate the performance for
this approach. Expanding from \eqnref{eqn:rec}, and substituting
$a_I(e^i_q,\epsilon)$ for the recall of each neighboring entity
$e^i_q$ for $e_q$, the recall for any query entity is:
\begin{eqnarray}
R(e_q,\epsilon,\delta) & = & a_I(e_q,\epsilon) + (1-a_I(e_q,\epsilon))\times
r_I(e_q,\delta)\times
\sum_{i=1}^{k} p^{e_q}_i a_I(e^i_q,\epsilon) \nonumber
\end{eqnarray}
Similarly, on substituting $a_A(e^i_q,e^j,\epsilon)$ in
\eqnref{eqn:imp} for the imprecision of each neighboring entity
$e^i_q$, we get the following expression for imprecision:
\begin{eqnarray}
I(e_q,e',\epsilon,\delta) & = & a_A(e_q,e',\epsilon) +
(1-a_A(e_q,e',\epsilon))\times r_A(e_q,e',\delta)\times
\sum_{i=1}^{k}\sum_{j=1}^{l} p^{e_q}_i p^{e'}_j
a_A(e^i_q,e^{'j},\epsilon) \nonumber
\end{eqnarray}
To appreciate more easily the implications of considering
first-order neighbors, we may assume that the attribute identification
probability and the attribute ambiguity probability are the same for
all the entities involved, i.e., $a_I(e,\epsilon) = a_I(\epsilon)$
and $a_A(e,e',\epsilon)=a_A(\epsilon)$. Then, using
$\sum_{i=1}^{k}p^e_i=1$ for any entity $e$, the expression for
recall simplifies to
\begin{eqnarray}
R(e_q,\epsilon,\delta) & = & a_I(\epsilon) + (1-a_I(\epsilon))\times
r_I(\delta)\times a_I(\epsilon) \nonumber \\ & = & a_I(\epsilon) [1
+(1-a_I(\epsilon))r_I(\delta)] \nonumber
\end{eqnarray}
Similarly, the expression for imprecision simplifies to
\begin{eqnarray}
I(e_q,e',\epsilon,\delta) & = & a_A(\epsilon)[1 + (1-a_A(\epsilon)) r_A(\delta)]
\nonumber
\end{eqnarray}

So we can see that attribute-clustering of the first level
neighbors potentially increases recall for any query entity $e_q$, but
imprecision goes up as well. However, when the balance between $r_A$
and $r_I$ is favorable, the increase in imprecision is insignificant
and much smaller than the corresponding increase in recall, so that
there is an overall performance improvement.

Can we do better than this? We can go a step further and consider
co-occurrence relations for resolving the {\em level-1} references
as well. So, instead of considering attribute-based resolution for
references in {\em level-1} as before, we perform collective
resolution for them. We consider all of their $\delta$-similar
references, which we call {\em level-2} references ($Rel^2(Q)$),
using A-expansion:
\begin{displaymath}
Rel^2(Q)=X_A(Rel^1(Q))
\end{displaymath}
Note that we have overloaded the A-expansion operator for a set $R$
of references: $X_A(R) = \bigcup_{r\in R} X_A(r.A)$. The {\em
level-3} references are the second order neighbors that co-occur
with {\em level-2} references. They are retrieved using H-expansion
on the {\em level-2} references:
\begin{displaymath}
Rel^3(Q)=X_H(Rel^2(Q))
\end{displaymath}
Finally, as with the {\em level-1} references earlier, we resolve the
{\em level-3} references using $\epsilon$-similarity of their
attributes alone.

In order to evaluate the impact on resolution accuracy for the query,
we unfold the recursions in \eqnref{eqn:rec} and \eqnref{eqn:imp} up
to two levels, and now substitute $a_I(e^i_q,\epsilon)$ for recall and
$a_A(e^i,e^j,\epsilon)$ for imprecision for the second order
neighbors. The trend in the expressions becomes clearly visible if we
assume, as before, that $a_I$ and $a_A$ is identical for all entities,
and, additionally, $r_I$ and $r_A$ are also the same, i.e.,
$r_I(e_1,e_2,\epsilon)=r_I(\epsilon)$ and
$r_A(e_1,e_2,\delta)=r_A(\delta)$. Then, we can work through a few
algebraic steps to get the following expressions for recall and
precision for any query entity $e_q$:
\begin{eqnarray}
R(e_q) & = & a_I[1 + (1-a_I)r_I + (1-a_I)^2 r_I^2] \\
I(e_q,e') & = & a_A[1 + (1-a_A)r_A + (1-a_A)^2 r_A^2]
\label{eqn:gp}
\end{eqnarray}

We can continue to unfold the recursion further and grow the
relevant set for the query. Formally, the expansion process
alternates between A-expansion and H-expansion:
\begin{displaymath}
\begin{array}{lcll}
Rel^{i}(Q) & = & X_A(Q)               & \mbox{ for }\ i=0 \nonumber \\
             &   & X_H(Rel^{i-1}(Q)) & \mbox{ for odd}\ i \nonumber \\
             &   & X_A(Rel^{i-1}(Q)) & \mbox{ for even}\ i
\end{array}
\end{displaymath}

As we proceed recursively and consider higher order co-occurrences for
the query, additional terms appear in the expressions for precision
and recall. But this does not imply that we need to continue this
process to arbitrary levels to get optimum benefit. Using our
simplifying assumptions about the attribute and relational
probabilities, the expressions for both recall and imprecision for
$n^{th}$ order co-occurrences turns out to be geometric progressions
with $n+1$ terms. The common ratio for the two geometric progressions
are $(1-a_I(\epsilon))r_I(\delta)$ and
$(1-a_A(\epsilon))r_A(\delta)$ respectively. Typically, both
of these ratios are significantly smaller than 1, and therefore
converge very quickly with increasing co-occurrence level. So the
improvement in resolution accuracy for the query $Q$ falls off quickly
with expansion depth,
and we can terminate the expansion process at
some cut-off depth $d^*$ without compromising on accuracy:
\begin{displaymath}
Rel(Q) = \bigcup_{i=0}^{d^*} Rel^i(Q)
\end{displaymath}

Of course, the assumptions about the attribute and relational
probabilities being entity-independent do not hold in practice, so
that the performance trends for increasing levels of co-occurrence
cannot be exactly captured by geometric progressions with a common
ratio for successive terms. But the converging trends for both of
them still hold in general, and the rate of convergence is still
determined by the four probabilities $a_I, a_A, r_I$ and $r_A$ for
the entities that are encountered during the expansion process.
Intuitively, smaller values for $r_I$ and $r_A$ indicate less
sensitivity to co-occurrences, and the convergence is quicker. On
the other hand, higher values of $a_I$ and $a_A$ mean that more
entities are resolved based on attributes alone --- correctly or
incorrectly --- and the impact of co-occurrence relations is
smaller. Therefore convergence is quicker for higher values of $a_I$
and $a_A$.

Apart from imposing a cutoff on the expansion depth, the size of the
relevant set can also be significantly reduced by restricting
attribute expansion beyond {\em level-0} to {\bf exact A-expansion}
$X^e_A(r)$. This only considers references with exactly the same
attribute as $r$ and disregards other $\delta$-similar references.
Interestingly, we can show that the restricted strategy that
alternates between exact A-expansion and H-expansion does not reduce
recall significantly.

\section{Adaptive Query Expansion} \label{sec:adaptive}

The limited depth query expansion strategy proposed in the previous
section is an effective approach that is able to answer queries
quickly and accurately for many domains. However, for some domains,
the size of the relevant set that is generated can be extremely
large even for small expansion depths, and as a result, the
retrieved references cannot be resolved at query-time. In this
section, we propose adaptive strategies based on estimating the
`ambiguity' of individual references that makes our algorithm even
more efficient while preserving accuracy.

The main reason behind this explosive growth of the relevant set
with increasing levels is that our query expansion strategy from the
previous section is unconstrained --- it treats all co-occurrences
as equally important for resolving any entity. It blindly expands
all references in the current relevant set, and also includes all
new references generated by an expansion operation. Given the
limited time to process a query, this approach is infeasible for
domains that have dense relationships. Our solution is to identify
the references that are likely to be the most helpful for resolving
the query, and to focus on only those references. To illustrate
using our example from \figref{fig:expansion}, observe that `Chen'
and `Li' are significantly more common or `ambiguous' names than
`Ansari' --- even different `W. Wang' entities are likely to have
collaborators named `Chen' or `Li'. Therefore, when h-expanding
$Rel^0(r_q)$ for `W. Wang', `Ansari' is more informative than `Chen'
or `Li'. Similarly, when n-expanding $Rel^1(r_q)$, we can choose not
to expand the name `A. Ansari' any further, since two `A. Ansari'
references are very likely to be coreferent. But we need more
evidence for the `Chen's and the `Li's.

To describe this formally, the ambiguity of a value $a$ for an
attribute $A$ is the probability that any two references $r_i$ and
$r_j$ in the database that have $r_i.A=r_j.A=a$ are {\it not}
coreferent: $Amb(a) = P(E(r_i)\neq E(r_j) \mid r_i.A=r_j.A=a)$. The
goal of adaptive expansion is to add less ambiguous references to
the relevant set and to expand the most ambiguous references
currently in the relevant set. We first define adaptive versions of
our two expansion operators treating the ambiguity estimation
process as a black-box, and then look at ways to estimate ambiguity
of references.

\subsection{Adaptive Expansion Operators}

The goal of adaptive expansion is to selectively choose the references
to expand from the current relevant set, and also the new references
that are included at every expansion step. For {\bf adaptive
hyper-edge expansion}, we set an upper-bound $h_{max}$ on the number
of new references that h-expansion at a particular level can
generate. Formally, we want \\$|X_H(Rel^i(Q))|$ $\leq$
$h_{max}|Rel^i(Q)|$. The value of $h_{max}$ may depend on depth $i$
but should be small enough to rule out full h-expansion of the
current relevant set. Then, given $h_{max}$, our strategy is to
choose the least ambiguous references from $X_H(Rel^i(Q))$, since
they provide the most informative evidence for resolving the
references in $Rel^i(Q)$. To achieve this, we sort the h-expanded
references in increasing order of ambiguity and select the first $k$
from them, where $k=h_{max}|Rel^i(Q)|$.
\begin{equation}
Rel^i_{adapt}(Q, h_{max}) = LeastAmb(k, X_H(Rel^{i-1}_{adapt}(Q)))
\end{equation}

The setting for {\bf adaptive attribute expansion} is very
similar. For some positive number $a_{max}$, exact a-expansion of
$Rel^i(Q)$ is allowed to include at most $a_{max}|Rel^i(Q)|$
references. Note that now the selection preference needs to be flipped
--- more ambiguous names need more evidence, so they are expanded
first. So we can sort $X^e_A(Rel^i(Q))$ in decreasing order of
ambiguity and select the first $k$ from the sorted list, where
$k=a_{max}|Rel^i(Q)|$. But this could potentially retrieve only
references for the most ambiguous name, totally ignoring references
with any other name. To avoid this, we choose the top $k$ ambiguous
references from $Rel^i(Q)$ {\it before} expansion, and then expand the
references so chosen.
\begin{eqnarray}
\label{eqn:relv-adapt}
Rel^i_{adapt}(Q, n_{max}) & = & X^e_A(MostAmb(k,
Rel^i_{adapt}(Q)))
\end{eqnarray}
Though this cannot directly control the number of new references
added, $\mu_r\times k$ is a reasonable estimate, where $\mu_r$ is the
average number of references per name.

\subsection{Ambiguity Estimation}
The adaptive expansion scheme proposed in this section is crucially
dependent on the estimates of name ambiguity. We now describe one
possible scheme that worked quite well. Recall that we want to
estimate the probability that two randomly picked references with
value $a$ for attribute $A$ correspond to different entities. For a
reference attribute $A_1$, denoted $R.A_1$, a naive estimate for the
ambiguity of a value of $n$ for the attribute is:
\begin{displaymath}
Amb(r.A_1) =
\frac{| \sigma_{R.A_1 = r.A_1} (R) | }{ | R |},
\end{displaymath}
where $|\sigma_{R.A_1 = r.A_1} (R) |$ denotes the number of references
with value $r.A_1$ for $A_1$. This estimate is clearly not good since
the number of references with a certain attribute value does not
always match the number of different entity labels for that attribute.
We can do much better if we have an additional attribute $A_2$. Given
$A_2$, the ambiguity for value of $A_1$ can be estimated as
\begin{displaymath}
Amb(r.A_1\mid
r.A_2) = \frac{| \delta (\pi_{R.A_2} (\sigma_{R.A_1 = r.A_1} (R) ))| }{ | R |},
\end{displaymath}
where $| \delta (\pi_{R.A_2} (\sigma_{R.A_1 = r.A_1} (R)) )|$ is the
number of distinct values observed for $A_2$ in references with
$R.A_1=r.A_1$. For example, we can estimate the ambiguity of a last
name by counting the number of different first names observed for it.
This provides a better estimate of the ambiguity of any value of an
attribute $A_1$, when $A_2$ is not correlated with $A_1$.  When
multiple such uncorrelated attributes $A_i$ are available for
references, this approach can be generalized to obtain better
ambiguity estimates.

\begin{figure}[t]
\begin{minipage}[c]{5in}
\begin{tabbing}
aaaaaaaaaaaa \= aaaaaa \= aaaaaaaa \= aaaaaa \=> \kill
   \>  Algorithm Query-time Resolve ($\mathcal R$.Name name)\\
1. \>  RSet = RelevantFrontier(name) \\
2. \>  RC-ER(RSet) \\
\\
   \> Algorithm FindRelevantRefs($\mathcal R$.Name name) \\
1. \> Initialize RSet to \{\} \\
5. \> Initialize depth to 0 \\
3. \> Initialize FrontierRefs to \{\} \\
4. \> While depth $<$ d* \\
5. \> \>    If depth is even or 0 \\
6. \> \> \>     R = $X_A$(FrontierRefs)   \\
7. \> \>    else \\
8. \> \> \>     R = $X_H$(FrontierRefs) \\
9. \> \>    FrontierRefs = R \\
10.\> \>    Add FrontierRefs to RSet \\
10.\> \>    Increment depth \\
11.\> Return RSet \\
\end{tabbing}
\normalsize
\end{minipage}
\caption{High-level description of the query-time entity resolution algorithm}
\label{fig:qt-algo}
\end{figure}

Putting everything together,
high-level pseudo code for the query-time entity resolution algorithm
is shown in \figref{fig:qt-algo}. The algorithm works in two stages --- first, it identifies the relevant set of references given an entity name as a query, and then it performs relational clustering on the extracted relevant references. The relevant references are extracted using the recursive process that we have already seen. The relevant references at any depth $i$ are obtained by expanding the relevant references at depth $i-1$, the expansion being dependent of whether it is an odd step or an even step. The actual expansion operator that is used may either be unconstrained or adaptive.

\section{Empirical Evaluation} \label{sec:exp}

For experimental evaluation of our query-time resolution strategies,
we used both real-world and synthetically generated datasets. First,
we describe our real datasets and the experiments performed on them
and then we move on to our experiments on synthetic data.

\subsection{Experiments on Real Data}
For real-world data, we used two citation datasets with very
different characteristics. The first dataset, {\bf arXiv}, contains
papers from high energy physics and was used in KDD Cup
2003\footnote{{\scriptsize
http://www.cs.cornell.edu/projects/kddcup/index.html}}. It has
58,515 references to 9,200 authors, contained in 29,555
publications. The number of author references per publication ranges
from 1 to 10 with an average of 1.90. \commentout{There are 12886
distinct names with the number of references per name ranging from 1
to 178, with an average of 6.53.} Our second dataset is the {\bf
Elsevier BioBase} database\footnote{{\scriptsize
http://help.sciencedirect.com/robo/projects/sdhelp/about\_biobase.htm}}
of publications from biology used in the recent IBM KDD-Challenge
competition.  It includes all publications under `Immunology and
Infectious Diseases' between years 1998 and 2001. This dataset
contains 156,156 publications with 831,991 author references. The
number of author references per publication is significantly higher
than arXiv and ranges from 1 to 100 (average 5.3). All names in this
database only have initials for first and middle names (if
available),
unlike arXiv,
which has both initialed and complete
names. The number of distinct names in BioBase is 303,693, with the
number of references for any name ranging from 1 to 193 (average
2.7). Unlike arXiv, BioBase includes keywords, topic classification,
language, country of correspondence and affiliation of the
corresponding author as attributes of each paper, all of which we use
as attributes for resolution in addition to author names. BioBase is
diverse in terms of these attributes, covering 20 languages, 136
countries, 1,282 topic classifications and 7,798 keywords.

For entity resolution queries in arXiv, we selected all ambiguous
names that correspond to more than one author entity. This gave us
75 queries, with the number of true entities for the selected names
varying from 2 to 11 (average 2.4). For BioBase, we selected as
queries the top 100 author names with the highest number of
references. The average number of references for each of these 100
names is 106, and the number of entities for the selected names
ranges from 1 to 100 (average 32), thereby providing a wide variety
of entity resolution settings over the queries.

\begin{figure*}[htb]
\centering
\subfigure{
\includegraphics[angle=-90, width=0.45\linewidth]{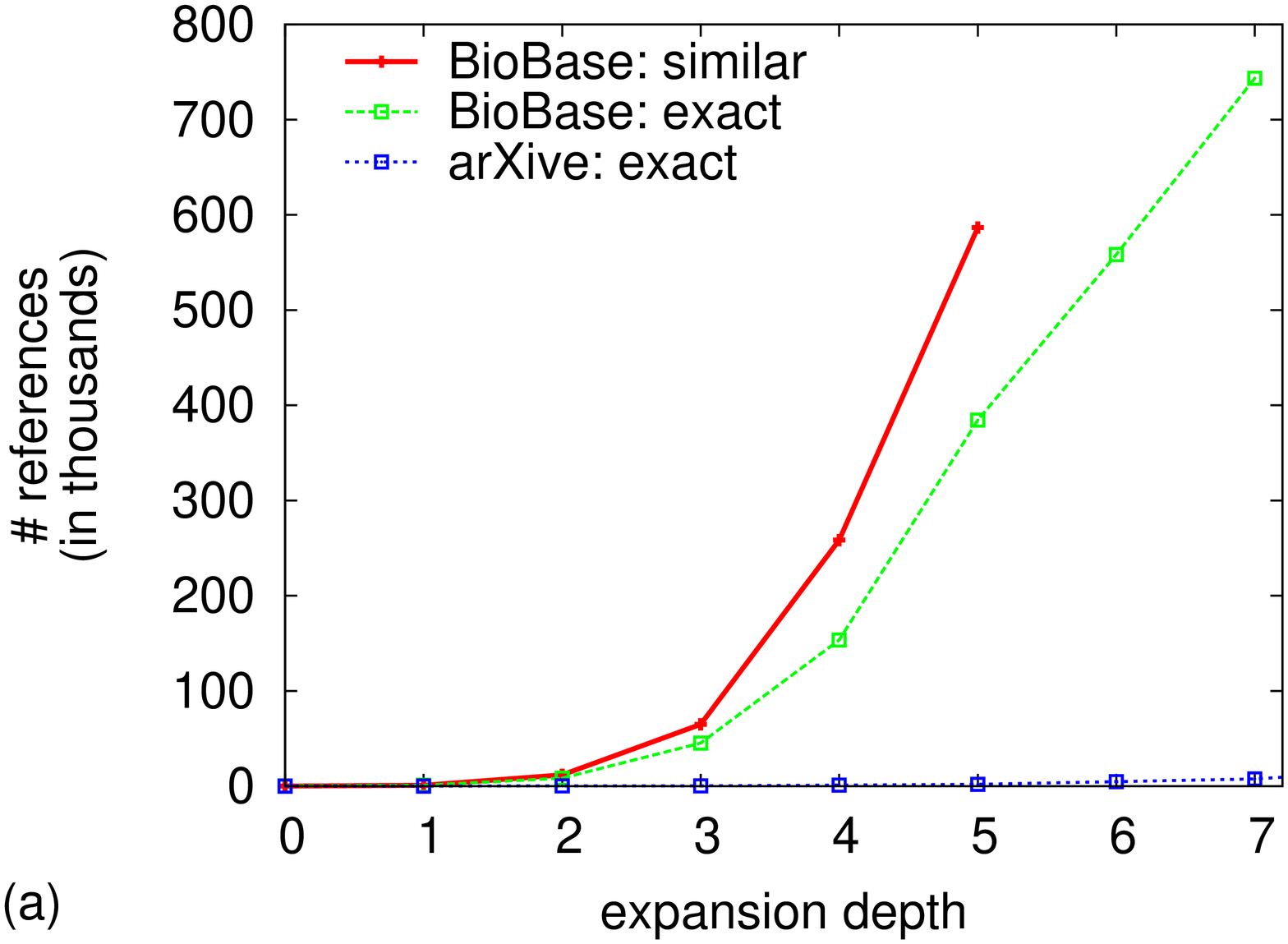}
}
\subfigure{
\includegraphics[angle=-90, width=0.45\linewidth]{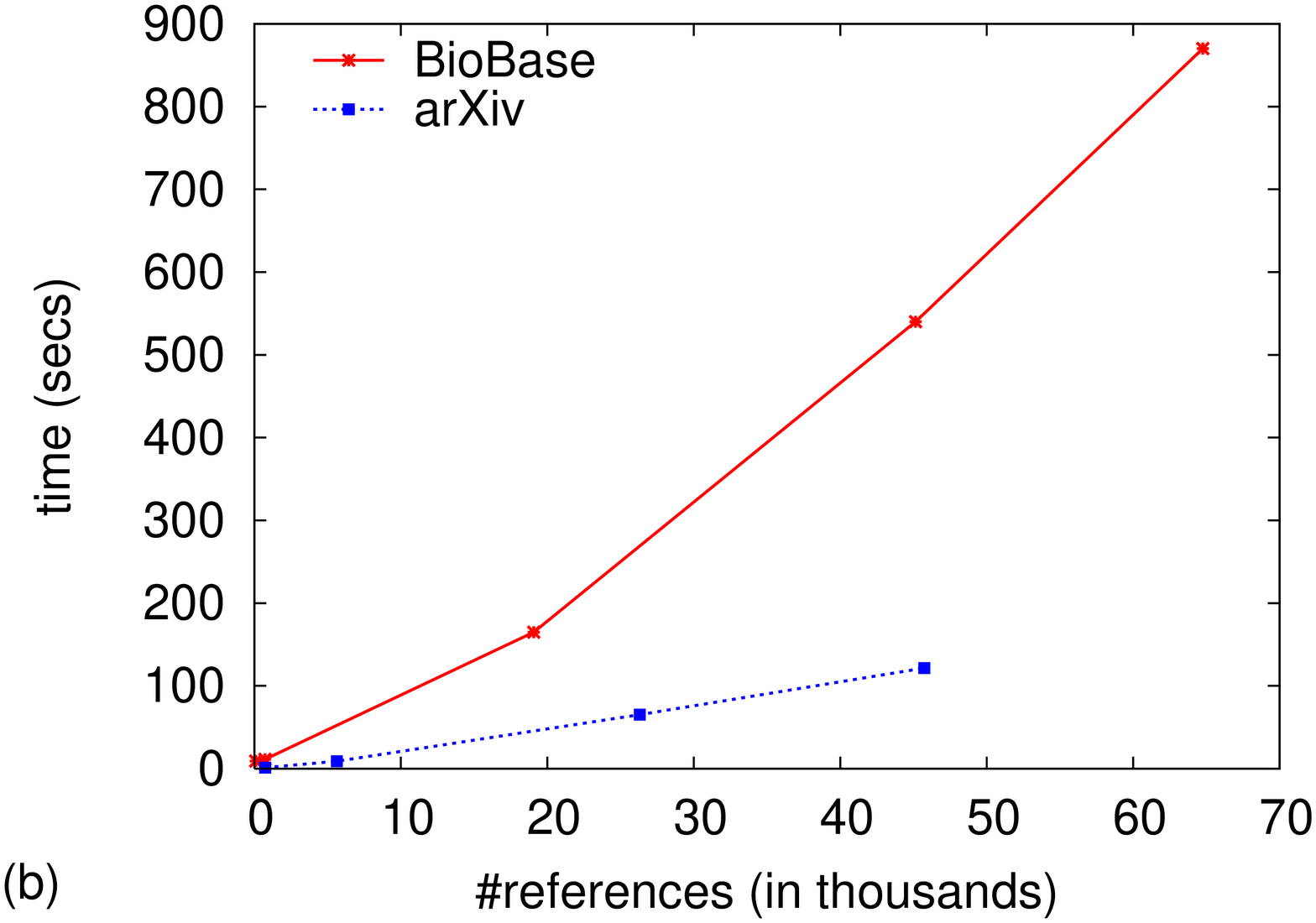}
}
\caption{(a) Size of the relevant set for increasing expansion depth for
sample queries in arXiv and BioBase (b) Execution time of RC-ER with
increasing number of references}
\label{plot:growth}
\end{figure*}

\subsubsection{Relevant Set Size Vs. Resolution Time}
We begin by exploring the growth rate of the relevant set for a
query over expansion depth in the two datasets.
\figref{plot:growth}(a) plots the size of the relevant set for a
sample query on the name `T. Lee' for arXiv and `M. Yamashita' for
BioBase. The growth rate for the arXiv query is moderate. The number
of references with name `T. Lee' is 7, which is the number of
relevant references at depth 0, and the size grows to 7,500 by depth
7. In contrast, for BioBase the plots clearly demonstrate the
exponential growth of the relevant references with depth for both
name expansion strategies. There are 84 relevant references at depth
0. When references are expanded using name similarity expansion,
there are 722 relevant references at depth 1, 65,000 at depth 3 and
more than 586,000 at depth 5. This is for a very restricted
similarity measure where two names are considered similar only if
their first initials match, and the last names have the same first
character and differ overall by at most 2 characters. A more liberal
measure would result in a significantly faster growth. We also
observe that for exact expansion, the growth is slower but we still
have 45,000 references at depth 3, 384,000 at depth 5 and 783,000 by
depth 7. It is interesting to note that the growth slows down beyond
depth 5; but this is because most of the references in the entire
dataset are already covered at that depth (BioBase has 831,991
references in total). The growth rates for these two examples from
arXiv and BioBase are typical for all of our queries in these two
datasets.

Next, in \figref{plot:growth}(b), we observe how the relational
clustering algorithm {\bf RC-ER} scales with increasing number of
references in the relevant set. All execution times are reported on
a Dell Precision 870 server with 3.2GHz Intel Xeon processor and 3GB
of memory. The plot shows that the algorithm scales well with
increasing references, but the gradient is different for the two
datasets. This is mainly due to the difference in the average number
of references per hyper-edge. This suggests that for arXiv, {\bf
RC-ER} is capable of handling the relevant sets generated using
unconstrained expansion. But for BioBase, it would require up to 600
secs for 40,000 references, and up to 900 secs for 65,000. So it is
clearly not possible to use {\bf RC-ER} with unconstrained expansion
for query-time resolution in BioBase even for depth 3.

\subsubsection{Entity Resolution Accuracy for Queries} In our next
experiment, we evaluate several algorithms for entity resolution
queries. We compare entity resolution accuracy of the pair-wise
co-reference decisions using the F1 measure (which is the harmonic
mean of precision and recall). For a fair comparison, we consider
the best F1 for each of these algorithms over all possible
thresholds for determining duplicates. For the algorithms, we
compare {\it attribute-based entity resolution} ({\bf A}), {\it
naive relational entity resolution} ({\bf NR}) that uses {\it
attributes} of related references, and our {\it relational
clustering algorithm for collective entity resolution} ({\bf RC-ER})
using unconstrained expansion up to depth 3. We also consider
transitive closures over the pair-wise decisions for the first two
approaches ({\bf A*} and {\bf NR*}). For attribute similarity, we
use the {\em Soft TF-IDF} with Jaro-Winkler similarity for names,
which has been shown to perform the best for name-based resolution
\cite{bilenko:is03}, and TF-IDF similarity for the other textual
attributes.

\begin{table}[t]
\caption{Average entity resolution accuracy (F1) for different
algorithms over 75 arXiv queries and 100 BioBase queries}
\label{table:performance}
\vskip 0.15in
\begin{center}
\begin{tabular}{|l|c|c|}
\hline
            & {\bf arXiv} & {\bf BioBase} \\
\hline
\hline
{\bf A}             & 0.721     & 0.701 \\
{\bf A*}        & 0.778     & 0.687 \\
{\bf NR}       & 0.956     & 0.710 \\
{\bf NR*}      & 0.952     & 0.753 \\
{\bf RC-ER Depth-1}     & 0.964     & 0.813 \\
{\bf RC-ER Depth-3}     & 0.970     & 0.820 \\
\hline
\end{tabular}
\end{center}
\vskip -0.1in
\end{table}

The average F1 scores over all queries are shown in
\tabref{table:performance} for each algorithm in the two datasets.
It shows that {\bf RC-ER} improves accuracy significantly over the
baselines. For example in BioBase, the improvement is $21\%$ over
{\bf A} and {\bf NR}, $25\%$ over {\bf A*} and $13\%$ over {\bf
NR*}. This demonstrates the potential benefits of collective
resolution for answering queries, and validates recent results in
the context of offline entity resolution
\shortcite{bhattacharya:dmkd04,bhattacharya:tkdd07,parag:kdd04-wkshp,dong:sigmod05,mccallum:nips04}. In our earlier work \shortcite{bhattacharya:tkdd07} we have demonstrated using extensive experiments on real and synthetic datasets how our relational clustering algorithm ({\bf RC-ER}) improves entity resolution performance over traditional baselines in the context of offline data cleaning, where the entire database is cleaned as a whole. The numbers in \tabref{table:performance} confirm that similar improvements can be obtained for localized resolution as well.
As predicted by our analysis, most of the accuracy improvement comes
from the depth-1 relevant references. For 56 out of the 100 BioBase
queries, accuracy does not improve beyond the depth-1 relevant
references. For the remaining 44 queries, the average improvement is
$2\%$. However, for 8 of the most ambiguous queries, accuracy
improves by more than $5\%$, the biggest improvement being as high
as $27\%$ (from 0.67 to 0.85 F1). Such instances are fewer for
arXiv, but the biggest improvement is $37.5\%$ (from 0.727 to 1.0).
On one hand, this shows that considering related records and
resolving them collectively leads to significant improvement in
accuracy. On the other hand, it also demonstrates that while there
are potential benefits to considering higher order neighbors, they
fall off quickly beyond depth 1. This also serves to validate our
analysis of collective query resolution in \secref{sec:rc-analysis}.

\commentout{
\begin{table}[t]
\caption{Average query processing time with unconstrained expansion}
\label{table:unconstrained} \vskip 0.15in
\begin{center}
\begin{tabular}{|l|cc|}
\hline
 & {\bf Avg \#references} & {\bf Time (in cpu secs)} \\
\hline \hline
{\bf arXiv} & 406 & 1.6 \\
{\bf BioBase} & 44,129 & 607  \\
\hline
\end{tabular}
\end{center}
\vskip -0.1in
\end{table}
}

\begin{table}[ht]
\caption{Average query processing time with unconstrained expansion}
\label{table:time}
\vskip 0.15in
\begin{center}
\begin{tabular}{|l|c|c|}
\hline
            & {\bf arXiv} & {\bf BioBase} \\
\hline
\hline
{\bf A}        & 0.41     & 9.35 \\
{\bf A*}       & 0.41     & 9.59 \\
{\bf NR}       & 0.43     & 28.54 \\
{\bf NR*}      & 0.428     & 28.69 \\
{\bf RC-ER Depth-1}     & 0.45     & 11.88 \\
{\bf RC-ER Depth-3}     & 1.36     & 606.98 \\
\hline
\end{tabular}
\end{center}
\vskip -0.1in
\end{table}

\begin{figure*}[t]
\centering
\subfigure[]{
\includegraphics[angle=-90, width=0.45\linewidth]{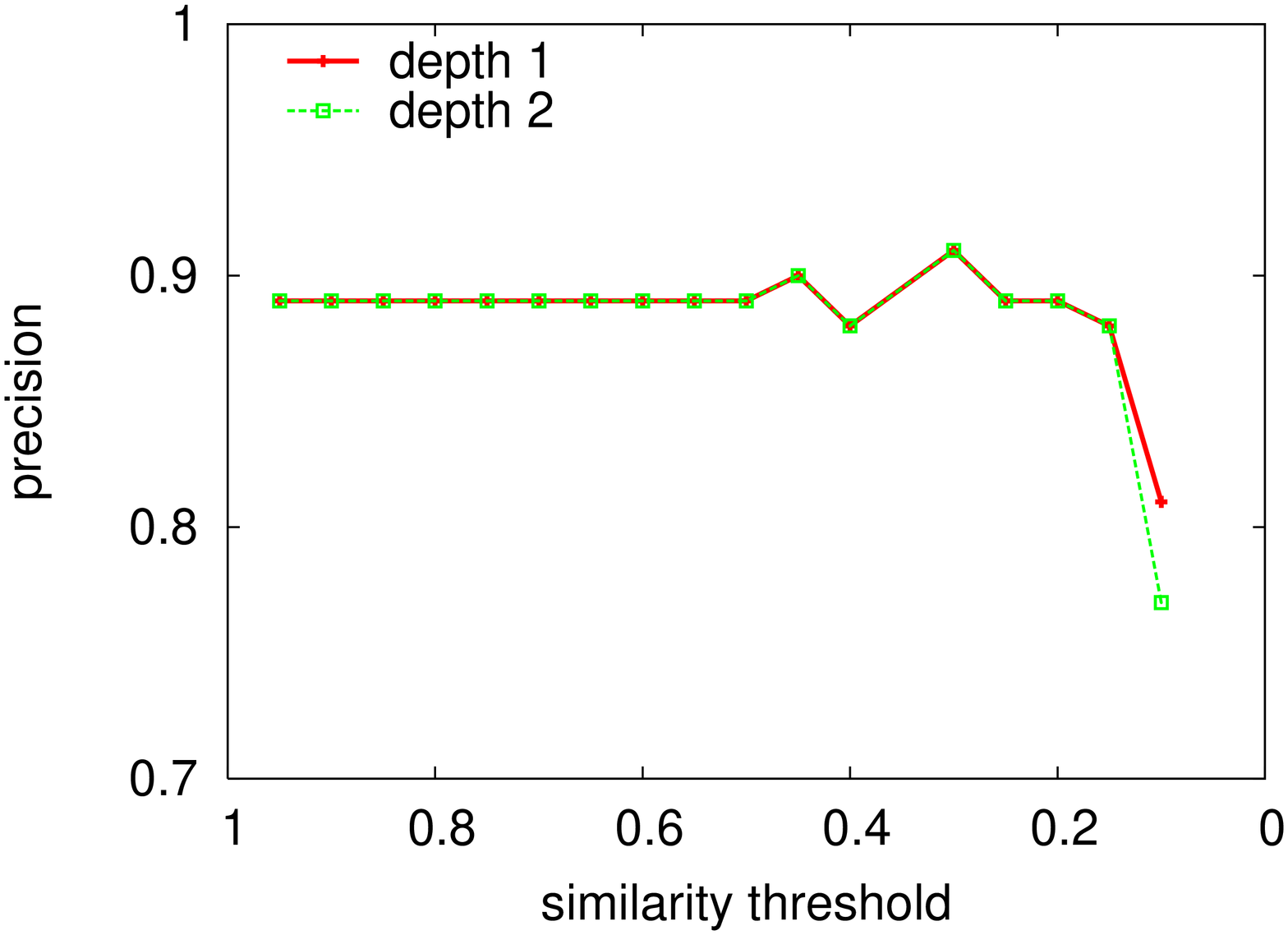}
}
\subfigure[]{
\includegraphics[angle=-90, width=0.45\linewidth]{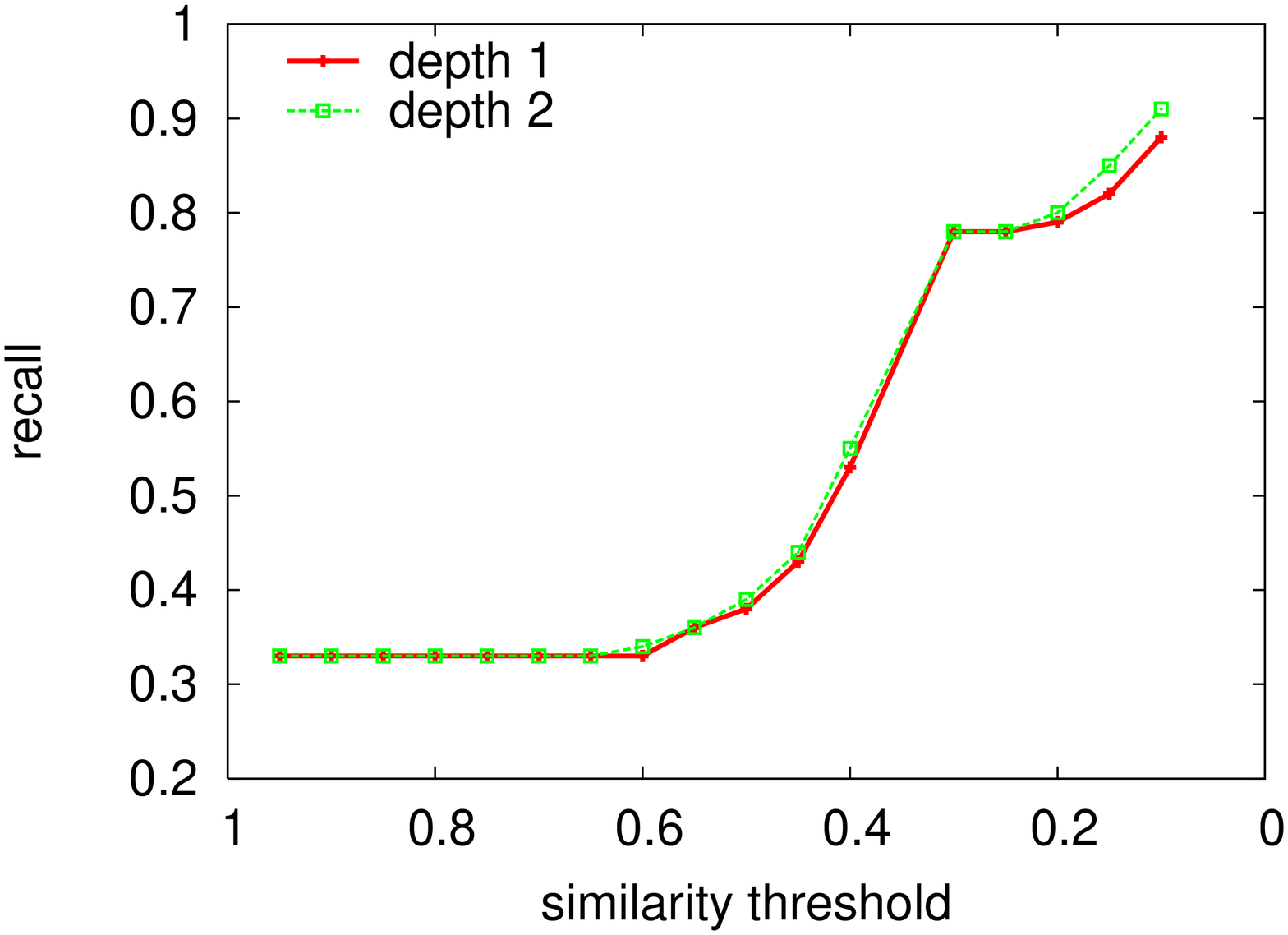}
} \\
\subfigure[]{
\includegraphics[angle=-90, width=0.45\linewidth]{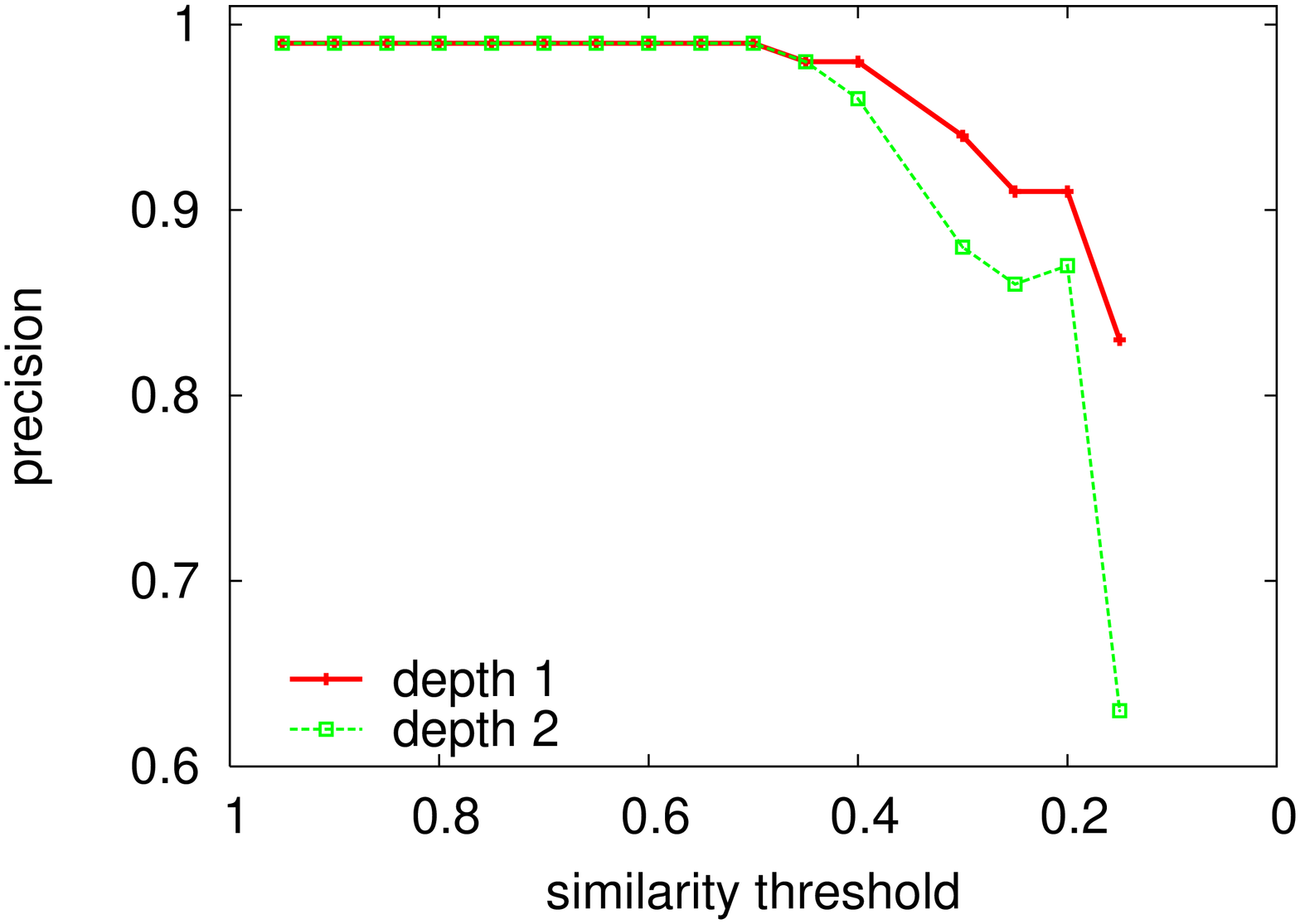}
}
\subfigure[]{
\includegraphics[angle=-90, width=0.45\linewidth]{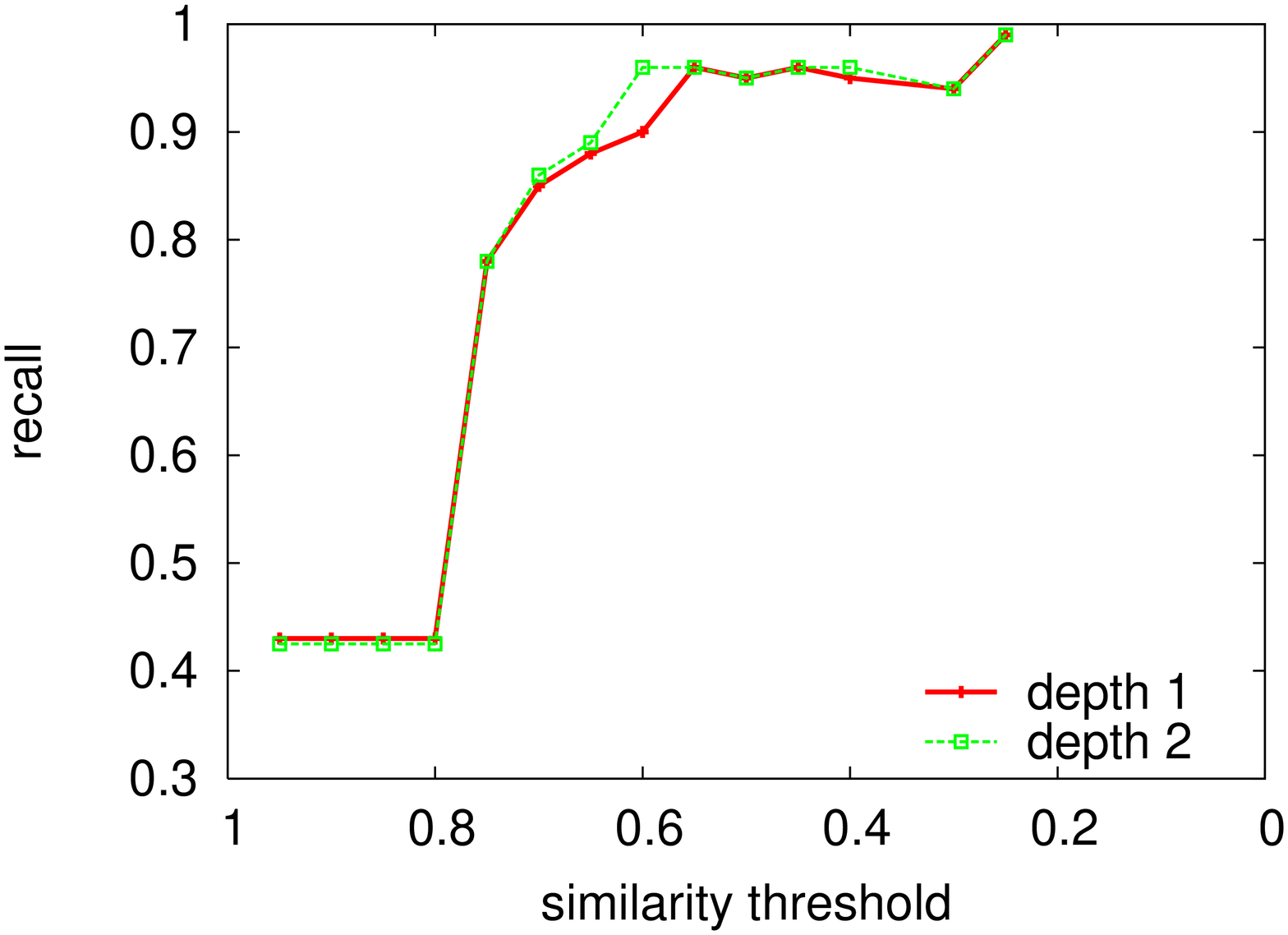}
} \\
\caption{Average precision and recall at different similarity thresholds for (a-b) BioBase and (c-d) arXiv}
\label{plot:gp}
\end{figure*}

The last two rows of \tabref{table:performance} show the converging nature of entity resolution performance with increasing depth. We verify this explicitly for precision and recall in \figref{plot:gp}. The top two plots show average precision and recall over BioBase queries at different similarity thresholds for {\bf RC-ER}. The bottom two plots show the same for arXiv. We can see that the precision curve at depth 1 coincides with or stays marginally above the precision curve at depth 3 for both BioBase and arXiv. The recall curves show the opposite trend --- recall marginally improves for depth 3. This is in agreement with our derived expressions for precision and recall for increasing depth in \eqnref{eqn:gp}. The difference in recall between depths 1 and 3 can be quantified as $a_I (1-a_I)^2 r_I^2$, and the difference in precision as $a_A (1-a_A)^2 r_A^2$. The explanation for the small difference between average precision and recall in these two plots is that both of these factors, when averaged over all queries, are significantly smaller than 1 for arXiv and BioBase. We will investigate this converging nature of performance in more detail by varying these structural properties in our experiments with synthetic data in \secref{subsec:synth}.

\subsubsection{Reducing Time with Adaptive Expansion}
 The first set of experiments show the effectiveness of
our two-phase query processing strategy in terms of entity
resolution performance. The challenge, as we have described earlier, is in obtaining these benefits in real-time. So, next, we focus on the time that is required
to process these queries in the two datasets using unconstrained
expansion up to depth 3. The results are shown in
\tabref{table:time}. For arXiv, the average processing time for depth-3 expansion
is 1.36 secs, with 406 relevant references on average. This shows that our two-phase strategy with
unconstrained expansion is a practical processing strategy for
entity resolution queries --- it resolves the query entities
accurately, and extremely quickly as well. However, for BioBase, the average number of references reached by depth 3 is more that 44,000, and the time taken to resolve them collectively
is more than 10 minutes. This is unacceptable for
answering queries, and next we focus on how the processing time is
improved using our proposed adaptive strategies. Note that the time taken for depth-1 expansion is around 12 secs, which is close to that for the attribute-based baseline ({\bf A}) and less than the time for the naive relational algorithm ({\bf NR}).

Since unconstrained expansion is effective for arXiv, we focus only
on BioBase for evaluating our adaptive strategies. For estimating
ambiguity of references, we use last names with first initial as the
secondary attribute. This
results
in very good estimates of
ambiguity --- the ambiguity estimate for a name is strongly
correlated (correlation coeff. 0.8) with the number of entities for
that name. First, we evaluate adaptive H-expansion. Since
H-expansion occurs first at depth 1, for each query, we construct
the relevant set with cutoff depth $d^*=1$, and use adaptive
H-expansion for depth 1. The expansion upper-bound $h_{max}$ is set
to 4. We compare three different adaptive H-expansion strategies:
(a) choosing the least ambiguous references, (b) choosing the most
ambiguous references and (c) random selection. Then, for each query,
we evaluate entity resolution accuracy using {\bf RC-ER} on the
relevant sets constructed using these three adaptive strategies. The
average accuracies for the three strategies over all 100 queries are
shown in the first column of \tabref{table:adaptive}. Least
ambiguous selection, which is the strategy that we propose, clearly
shows the biggest improvement and most ambiguous the smallest, while
random selection is in between. Notably, even without many of the
depth-1 references, all of them improve accuracy over {\bf NR*} by
virtue of collective resolution.

We perform a similar set of experiments for evaluating adaptive
attribute expansion. Recall that depth 2 is the lowest depth where
adaptive attribute expansion is performed. So for each query, we
construct the relevant set with $d^*=3$ using adaptive A-expansion
at depth 1 and unconstrained  H-expansion at depths 1 and 3. The
expansion upper-bound $a_{max}$ is set to $0.2$, so that on average
1 out of 5 names are expanded. Again, we compare three strategies:
(a) expanding the least ambiguous names, (b) expanding the most
ambiguous names and (c) random expansion. The average accuracies for
the three schemes over all 100 queries are listed in the second
column of \tabref{table:adaptive}. The experiment with adaptive
A-expansion does not bring out the difference between the three
schemes as clearly as adaptive H-expansion. This is because we are
comparing A-expansion at depth 2 and, on average, not much
improvement can be obtained beyond depth 1 because of a ceiling
effect. But it shows that almost all the benefit up to depth 3 comes
from our proposed strategy of expanding the most ambiguous names.
\begin{table}[t]
\caption{Avg. resolution accuracy in F1 with different adaptive expansion strategies}
\label{table:adaptive}
\vskip 0.15in
\begin{center}
\begin{tabular}{|l|cc|}
\hline
 & {\bf H-expansion} & {\bf A-expansion} \\
\hline
\hline
{\bf Least Ambiguous} & 0.790 & 0.815 \\
{\bf Most Ambiguous} & 0.761 & 0.821  \\
{\bf Random} & 0.770 & 0.820 \\
\hline
\end{tabular}
\end{center}
\vskip -0.1in
\end{table}

The above two experiments demonstrate the effectiveness of the two
adaptive expansion schemes in isolation. Now, we look at the results
when we use them together. For each of the 100 queries, we construct
the relevant set $Rel(r_q)$ with $d^*=3$ using adaptive H-expansion
and adaptive exact A-expansion. Since most of the improvement from
collective resolution comes from depth-1 references, we consider two
different experiments. In the first experiment ({\bf AX-2}), we use
adaptive expansion only at depths 2 and beyond, and unconstrained
H-expansion at depth 1. In the second experiment ({\bf AX-1}), we
use adaptive H-expansion even at depth 1, with $h_{max}=6$.  For
both of them, we use adaptive expansion at higher depths 2 and 3
with parameters $h_{max}=3$ at 3 and $a_{max}=0.2$ at 2.

\begin{table}[ht]
\caption{Comparison between unconstrained
and adaptive expansion for BioBase}
\label{table:adapt}
\vskip 0.15in
\begin{center}
\begin{tabular}{|l|ccc|}
\hline
            & {\bf Unconstrained}   & {\bf AX-2}    & {\bf AX-1}    \\
\hline
\hline
{\bf relevant-set size}     & 44,129.5  & 5,510.52  & 3,743.52 \\
{\bf time (cpu secs)}       & 606.98    & 43.44     & 31.28 \\
{\bf accuracy (F1)}     & 0.821     & 0.818     & 0.820 \\
\hline
\end{tabular}
\end{center}
\vskip -0.1in
\end{table}

In \tabref{table:adapt}, we compare the two adaptive schemes against
unconstrained expansion with $d^*=3$ over all queries. Clearly,
accuracy remains almost unaffected for both schemes. First, we note
that {\bf AX-2} matches the accuracy of unconstrained expansion, and
shows almost the same improvement over depth 1. This accuracy is
achieved even though it uses adaptive expansion that expands a small
fraction of $Rel^1(Q)$, and thereby reduces the average size of the
relevant set from 44,000 to 5,500. More significantly, {\bf AX-1}
also matches this improvement even without including many depth-1
references. This reduction in the size of the relevant set has an
immense impact on the query processing time. The average processing
time drops from more than 600 secs for unconstrained expansion to 43
secs for {\bf AX-2}, and further to just 31 secs for {\bf AX-1},
thus making it possible to use collective entity resolution for
query-time resolution.

\subsubsection{Adaptive Depth Selection}
 As a further improvement, we investigate if processing
time can be reduced by setting the expansion depth $d^*$ adaptively,
depending on the ambiguity of the query name, as compared to a fixed
$d^*$ for all queries. In a simple setup, we set $d^*$ to 1 for
queries where the number of different first initials for a last name
is less than 10 (out of 26), and explore depth 2 only for more
ambiguous queries. This reduces expansion depth from 2 to 1 for 18
out of the 100 queries. As a result, the average processing time for
these queries is reduced by 35\% to 11.5 secs from 17.7 secs with no
reduction in accuracy. For three of these queries,
the original
processing time at depth 2 is greater than 30 secs. In these
preliminary experiments, we only evaluated our original set of 100
queries that are inherently ambiguous. In a more general setting,
where a bigger fraction of queries have lower ambiguity, the impact
is expected to be even more significant.

\subsection{Experiments using Synthetic Data}
\label{subsec:synth}

In addition to experiments on real datasets, we performed
experiments on synthetically generated data. This enables us to
reason beyond specific datasets, and also to empirically verify our
performance analysis for relational clustering in general, and more
specifically for entity resolution queries. We have designed a
generator for synthetic data \shortcite{bhattacharya:tkdd07} that allows
us to control different properties of the underlying entities and
the relations between them, and also of the observed co-occurrence
relationships between the entity references. Among other properties,
we can control the number of
entities,
the average number of
neighbor entities per entity, and the number and average size of
observed co-occurrences. Additionally, we can control the ambiguity
of entity attributes, and the number of ambiguous relationships
between entities. We present an overview of the synthetic data
generation process in Appendix A.

\begin{figure*}[t]
\centering
\subfigure{
\includegraphics[angle=0, width=0.45\linewidth]{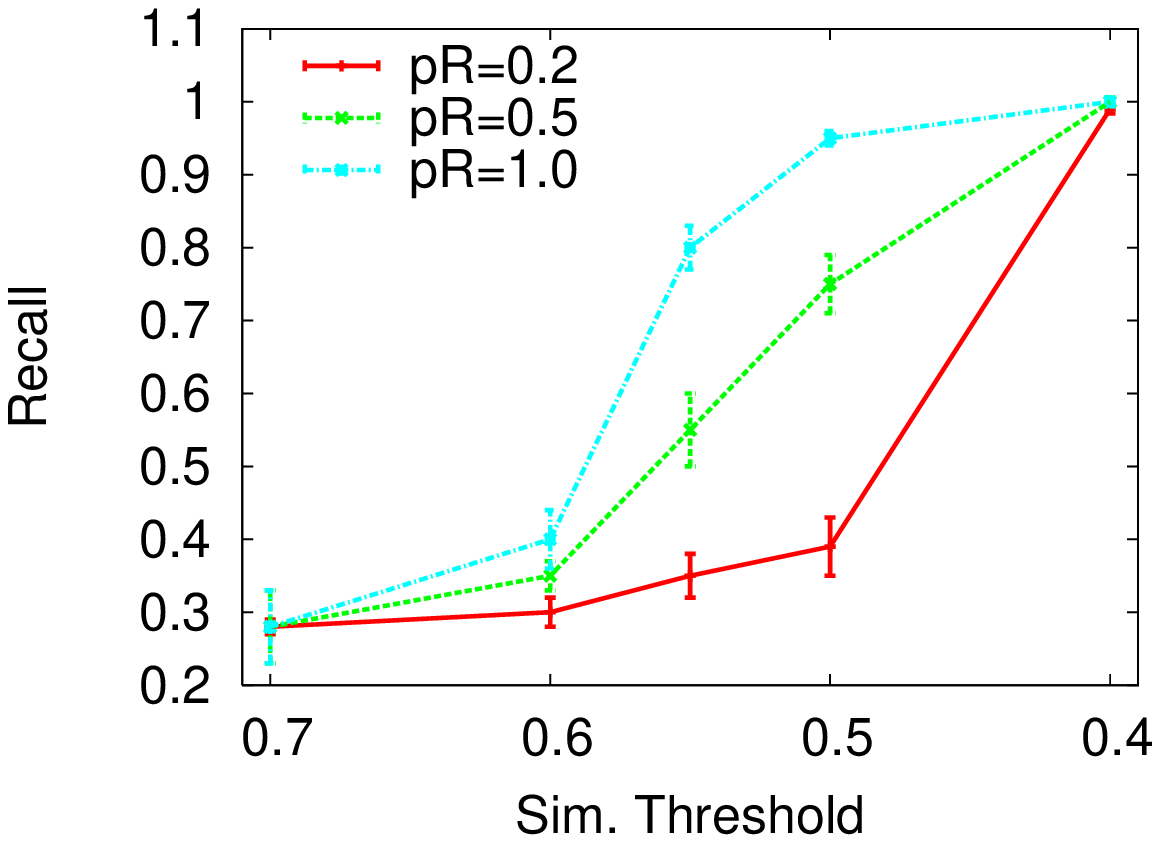}
}
\subfigure{
\includegraphics[angle=0, width=0.45\linewidth]{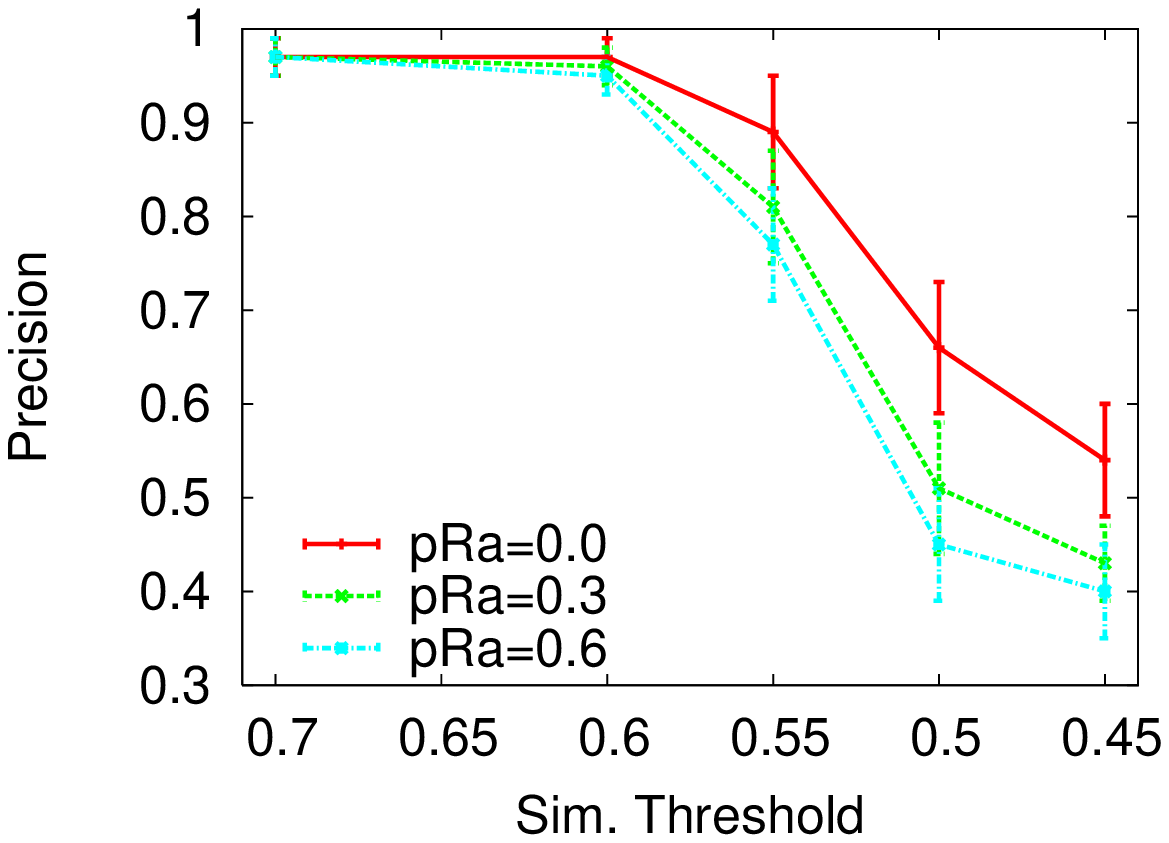}
} \caption{Effect of (a) identifying relations on recall and (b)
ambiguous relations on precision for collective clustering. Error
bars show standard deviation.} \label{plot:1}
\end{figure*}

We have performed a number of different experiments on synthetic
data. In the first set of experiments, we investigate the influence
of identifying relationships on collective resolution using
relational clustering. We generate 500 co-occurrence relations from
the same 100 entities and 200 entity-entity relationships, using
varying probability of co-occurrences $p^R=\{0.2,0.5,1.0\}$ in the
data. The probability of ambiguous relationships is held fixed, so
that higher $p^R$ translates to higher probability of identifying
co-occurrences in the data. \figref{plot:1}(a) shows recall at
different similarity thresholds for three different co-occurrence
probabilities. The results confirm that recall increases
progressively with more identifying relationships at all thresholds.
The curves for $p^R=0.5$ and $p^R=1.0$ flatten out only when no
further recall is achievable.

Next, we observe the effect of ambiguous relations on
the precision of
collective resolution using relational clustering. We add 200 binary
relationships between 100 entities in three stages with increasing
ambiguous relationship probability ($p^R_a=\{0,0.3,0.6$\}). Then we
perform collective resolution on 500 co-occurrence relations
generated from each of these three settings. In \figref{plot:1}(b)
we plot precision at different similarity threshold for three
different values of $p^R_a$. The plots confirm the progressive
decrease in precision for all thresholds with higher $p^R_a$. For
both experiments, the results are averaged over 200 different runs.

\begin{figure*}[t]
\centering
\subfigure{
\includegraphics[angle=0, width=0.45\linewidth]{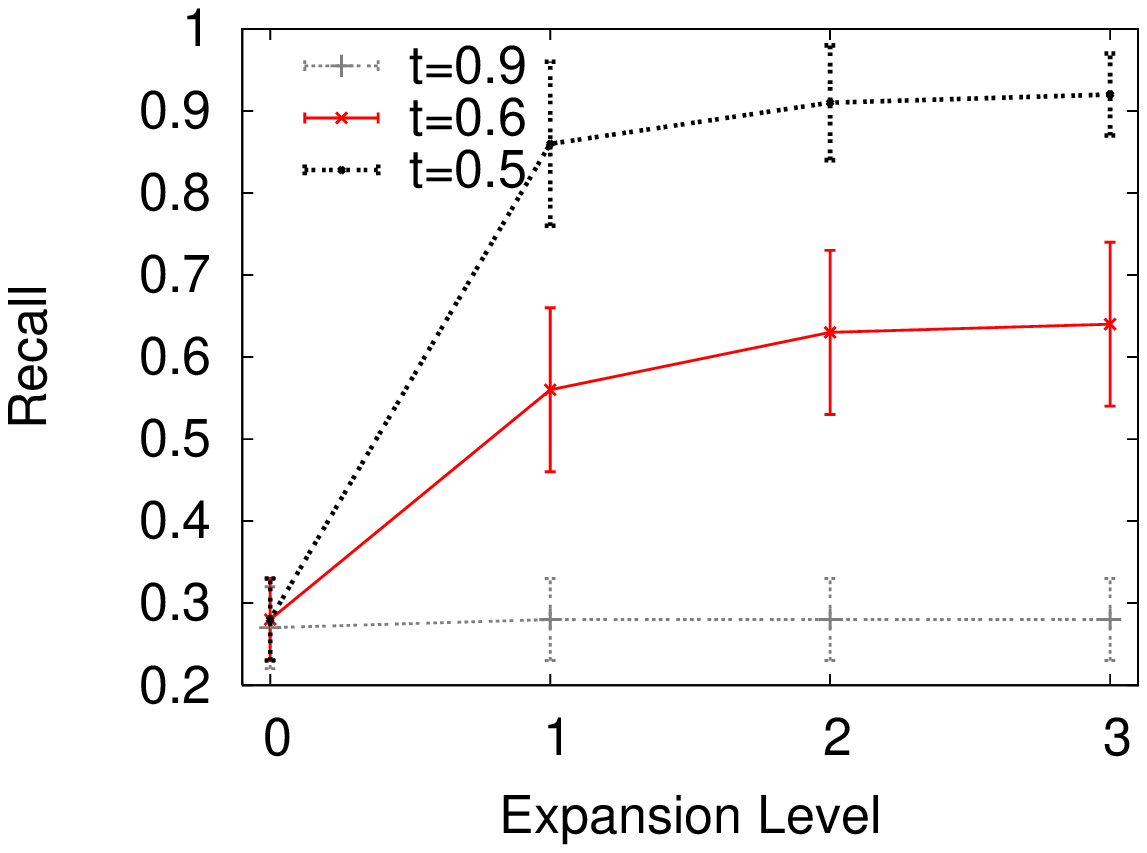}
}
\subfigure{
\includegraphics[angle=0, width=0.45\linewidth]{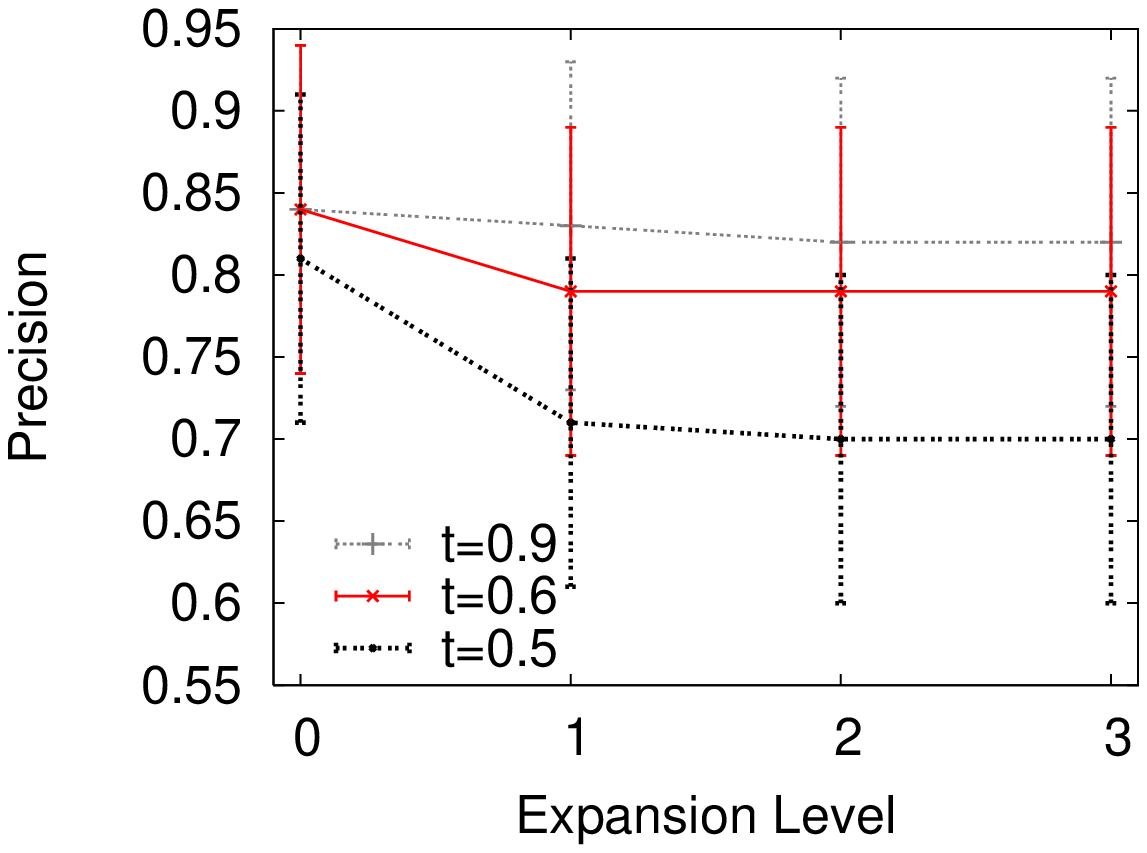}
} \caption{Change in (a) precision and (b) recall for increasing
expansion levels used for collective clustering. Error bars show
standard deviation.} \label{plot:2}
\end{figure*}

Next,
we evaluate collective resolution for queries. Recall that
the last two rows in \tabref{table:performance} clearly demonstrate
the converging nature of performance over increasing expansion
levels for queries on real datasets. We ran further experiments on
synthetic data to verify this trend. In each run, we generated 2,500
co-occurrence relations from 500 entities having an average of 2
neighbors per entity. Then we performed localized collective
clustering in each case, using as query the most ambiguous attribute
value (that corresponds to the highest number of underlying
entities). In \figref{plot:2}(c) and (d), we show how recall and
precision change with increasing expansion level for a query. Recall
improves with increasing expansion level, while precision decreases
overall, as is predicted by our analysis. Importantly, recall
increases at a significantly faster rate than that for the decrease
in precision. In general, the rate of increase/decrease depends on
the structural properties of the data, as we have shown in our
analysis. In other experiments, we have seen different rates of
change, but the overall trend remains the same. Our analysis also
showed that precision and recall converge quickly over increasing
expansion levels. This too is confirmed by the two plots where the
curves flatten out by level 3.

\subsection{Current Limitations}
Finally, we discuss two of the current limitations of our collective
entity resolution approach. Recall that the similarity measure in
Eqn. \ref{eqn:alpha} involves a weighting parameter $\alpha$ for
combining attribute and relational similarity.  For all of our
experiments, we report the best accuracy over all values of $\alpha$
for each query. Selecting the optimal value of $\alpha$ for each
query is an unresolved issue. However, our experiments reveal that
even a fixed $\alpha$ ($\alpha=0.5$) for all queries brings
significant improvements over the baselines.

The second issue is the determination of the termination threshold
for {\bf RC-ER}. Note that this is an issue for all of the baselines
as well, and here we report best accuracy over all thresholds. This
is an area of ongoing research. Preliminary experiments have shown
that the best threshold is often query specific --- setting the
threshold depending on the ambiguity of the query results in
significantly better accuracy than a fixed threshold for all
queries. For an empirical evaluation, we cleaned the entire arXiv
dataset offline by running {\bf RC-ER} on all its references
together, and terminated at the threshold that maximizes resolution
accuracy over all references. This results in an overall accuracy
(F1) of $0.98$. However, the average accuracy measured over the 75
queries in our test set is only $0.87$. In comparison, the best obtainable accuracy when resolving the queries individually
each with a different threshold is $0.97$. This suggests that there
may be potential benefits to localized cleaning over its global
counterpart in the offline setting.

\section{Related Work}
\label{sec:rw}

The entity resolution problem has been studied in many different
areas under different names --- deduplication, record linkage,
co-reference resolution, reference reconciliation, object
consolidation, etc. Much of the work has focused on traditional
attribute-based entity resolution. Extensive research has been done
on defining approximate string similarity measures
\cite{monge:kdd96,navarro:cs01,bilenko:is03,chaudhuri:sigmod03} that
may be used for unsupervised entity resolution. The other approach
uses adaptive supervised algorithms that learn similarity measures
from labeled data \cite{tejada:isj01,bilenko:kdd03}.

Resolving entities optimally is known to be computationally hard
even when only attributes are considered \cite{cohen:kdd00}.
Therefore, efficiency has received a lot of attention in
attribute-based data cleaning. The goal essentially is to avoid
irrelevant and expensive attribute similarity computations using a
`blocking' approach without affecting accuracy significantly
\shortcite{hernandez:sigmod95,monge:dmkd97,mccallum:kdd00}. The
merge/purge problem was posed by
\shortciteA{hernandez:sigmod95} with efficient schemes to retrieve
potential duplicates without resorting to quadratic complexity. They
use a `sorted neighborhood method' where an appropriate key is
chosen for matching. Records are then sorted or grouped according to
that key and potential matches are identified using a sliding window
technique. However, some keys may be badly distorted so that their
matches cannot be spanned by the window and such cases will not be
retrieved. The solution they propose is a multi-pass method over
different keys and then merging the results using transitive
closure. \shortciteA{monge:dmkd97} combine the union find
algorithm with a priority queue look-up to find connected components
in an undirected graph. \shortciteA{mccallum:kdd00}
propose the use of canopies to first partition the data into
overlapping clusters using a cheap distance metric and then use a
more accurate and expensive distance metric for those data pairs
that lie within the same canopy.
\shortciteA{chaudhuri:sigmod03} use an error tolerant index for data
warehousing applications for probabilistically looking up a small
set of candidate reference tuples for matching against an incoming
tuple. This is considered `probabilistically safe' since the closest
tuples in the database will be retrieved with high probability. This
is also efficient since only a small number of matches needs to be
performed. Swoosh \cite{benjelloun:tr05} has recently been proposed
as a generic entity resolution framework that considers resolving
and merging duplicates as a database operator and the goal is to
minimize the number of record-level and feature-level operations. An
alternative approach is to reduce the complexity of individual
similarity computations. \citeA{gravano:icde03}
propose a sampling approach to quickly compute cosine similarity
between tuples for fast text-joins within an SQL framework. All of these approaches enable efficient data cleaning when only attributes of references are considered.

Many recently proposed approaches take relations into account for
data integration
\shortcite{ananthakrishna:vldb02,bhattacharya:dmkd04,bhattacharya:kdd05-wkshp,kalashnikov:sdm05,dong:sigmod05}.
\shortciteA{ananthakrishna:vldb02} introduce
relational deduplication in data warehouse applications where there
is a dimensional hierarchy over the relations.
\shortciteA{kalashnikov:sdm05} enhance attribute similarity between an
ambiguous reference and the many entity choices for it with
relationship analysis between the entities, like affiliation and
co-authorship. In earlier work, we have proposed different measures
for relational similarity and a relational clustering algorithm for
collective entity resolution using relationships
\shortcite{bhattacharya:dmkd04,bhattacharya:tkdd07}.
\shortciteA{dong:sigmod05} collectively resolve entities of multiple types
by propagating relational evidences in a dependency graph, and
demonstrate the benefits of collective resolution in real datasets.
\citeA{long:icml06} have proposed a model for general multi-type relational clustering, though it has not been applied specifically for entity resolution. They perform collective factorization over related matrices using spectral methods to identify the cluster space that minimizes distortion over relationships and individual features at the same time.  All of these approaches that make use of relationships either for
entity matching (where the domain entities are known) or entity
resolution (where the underlying entities also need to be
discovered) have been shown to increase performance significantly
over the attribute-based solutions for the same problems. However,
the price they pay is in terms of computational complexity that
increases due to a couple of different reasons. Firstly, the number
of potential matches
increases
when relationships are considered and
individual similarity computations also become more expensive.
Secondly, collective resolution using relationships necessitates
iterative solutions that make multiple passes over the data. While
some of these approaches have still been shown to be scalable in
practice, they cannot be employed for query-time cleaning in a
straight-forward manner.

The idea of multi-relational clustering also comes up in the Inductive Logic Programming (ILP) literature. \citeA{emde:icml96} have used multi-relational similarity for instance-based classification of representations in first order logic. They define the similarity of two objects, e.g., of two people, as a combination of the similarity of their attribute values, such as their age, weight, etc., and the similarity of the objects that they are related to, such as the companies they work for. This is similar to the naive relational similarity that we discussed earlier, except that the similarity of the connected objects is also defined recursively in terms of their connected objects.
\citeA{kirsten:ilp98} have used this recursive relational similarity measure for agglomerative clustering of first order representations. While recursive comparison of neighbors is shown to be effective in terms of accuracy of results, the computational challenge is again a major drawback.

Probabilistic approaches that cast entity resolution as a
classification problem have been extensively studied. The groundwork
was done by \citeA{fellegi:jasa69}. Others
\cite{winkler:tr02,ravikumar:uai04} have more recently built upon
this work. Adaptive machine learning approaches have been proposed
for data integration \cite{sarawagi:kdd02,tejada:isj01}, where
active learning requires the user to label informative examples.
Probabilistic models that use relationships for collective entity
resolution have been applied to named entity recognition and
citation matching
\shortcite{pasula:nips03,mccallum:nips04,li:aimag05,parag:kdd04-wkshp}.
These probabilistic approaches are superior to similarity-based
clustering algorithms in that they associate a degree of confidence
with every decision, and learned models provide valuable insight
into the domain. However, probabilistic inference for collective entity resolution is not known to be scalable in practice, particularly when relationships are also
considered. These approaches have mostly been shown to work for
small datasets, and are significantly slower than their clustering
counterparts.

Little work has been done in the literature for query-centric
cleaning or relational approaches for answering queries, where
execution time is as important as accuracy of resolution. Approaches
have been proposed for localized evaluation of Bayesian networks
\cite{draper:uai94}, but not for clustering problems. Recently,
\citeA{chandel:icde06} have addressed efficiency
issues in computing top-k entity matches against a dictionary in the
context of entity extraction from unstructured documents. They
process top-k searches in batches where speed-up is achieved by
sharing computation between different searches.
\citeA{fuxman:sigmod05} motivate the problem of answering queries
over databases that violate integrity constraints and address
scalability issues in resolving inconsistencies dynamically at
query-time. However, the relational aspect of the problem, which is
the major scalability issue that we address, does not come up in any
of these settings. In our earlier work on relational clustering\shortcite{bhattacharya:tkdd07},
we used the idea of `relevant references' for experimental evaluation on the BioBase dataset. As we have also discussed here, this dataset has entity labels only for the 100 most frequent names. Therefore, instead of running collective resolution over
the entire BioBase dataset, we evaluated the 100 names separately, using only the `relevant references' in each case. The relevant references were the ones directly connected to references having the names of interest. The concept of focused cleaning, the performance analysis of relational clustering, the expand-resolve strategy and, most importantly, the idea of adaptive expansion for query-time resolution were not addressed in that paper.

One of the first papers to make use of relational features for classification problem was by \citeA{chakrabarti:sigmod98}. They showed that for the problem of classifying hyper-linked documents, naive use of relationships can hurt performance. Specifically, if key terms from neighboring documents are thrown into the document whose topic is to be classified, classification accuracy degrades instead of improving. The parallel in our scenario of clustering using relationships is that the naive relational model ({\bf NR}) may perform worse than the attribute model ({\bf A}) in the presence of highly ambiguous relationships. \shortciteA{chakrabarti:sigmod98} showed that relationships can however be used for improved classification when the {\em topic labels} of the neighboring documents are used as evidence instead of naively considering the terms that they contain. In our earlier work \shortcite{bhattacharya:dmkd04,bhattacharya:tkdd07}, we have shown similar results for collective clustering using relationships, where the cluster labels of neighboring labels lead to improved clustering performance compared to naive relational and attribute-based clustering. The interesting result that we have shown in this paper both in theory and empirically is that even collective use of relationships can hurt clustering accuracy compared to attribute-based clustering. This happens when relationships between references are dense and ambiguous, and errors that propagate over relationships exceed the identifying evidence that they provide.

\section{Conclusions} \label{sec:concl}

In this paper, we have motivated the problem of query-time entity
resolution for accessing unresolved third-party databases. For
answering entity resolution queries, we have addressed the
challenges of using collective approaches, which have recently shown
significant performance improvements over traditional baselines in
the offline setting. The first hurdle for collective resolution
arises from the interdependent nature of its resolution decisions.
We first formally analyzed the recursive nature of this dependency,
and showed that the precision and recall for individual entities
grow in a geometric progression as increasing levels of neighbors
are considered and collectively resolved. We then proposed a
two-stage `expand and resolve' strategy for answering queries based
on this analysis, using two novel expansion operators. We showed
using our analysis that it is sufficient to consider neighbors up to
small expansion depths, since resolution accuracy for the query
converges quickly with increasing expansion level. The second
challenge for answering queries is that the computation has to be
quick. To achieve this, we improved on our unconstrained expansion
strategy to propose an adaptive algorithm, which dramatically
reduces the size of the relevant references --- and, as a result,
the processing time --- by identifying the most informative
references for any query. We demonstrated using experiments on two
real datasets that our strategies enable collective resolution at
query-time, without compromising on accuracy. We additionally
performed various experiments on synthetically generated data over a
wide range of settings to verify the trends predicted by our
analysis. In summary, we have addressed and motivated a critical
data integration and retrieval problem, proposed algorithms for
solving it accurately and efficiently, provided a theoretical
analysis to validate our approach and explain why it works, and,
finally, shown experimental results on multiple real-world and
synthetically generated datasets to demonstrate that it works
extremely well in practice. While we have presented results for
bibliographic data, the techniques are applicable in other
relational domains.

While we have shown the dramatic reduction in query processing time that comes with adaptive expansion, more research is necessary to be able to answer entity resolution queries
on the order of milli-seconds, as may be demanded in many scenarios. Interesting directions of future research include exploring stronger coupling between the extraction and
resolution phases of query processing, where the expansion happens ``on-demand" only when the resolution process finds the residual ambiguity to be high and requires additional evidence for taking further decisions. This would directly address the problem of determining the expansion depth. While we have reported some preliminary experiments in this paper, more work needs to be done on adaptive depth determination depending on ambiguity. In the same context, we may imagine ``soft" thresholds for adaptive expansion, where the expansion operator automatically determines the number of hyper-edges or names to be expanded so that the residual ambiguity falls below some specified level. Other interesting extensions include caching of intermediate resolutions, where the related resolutions performed for any query are stored and retrieved as and when required for answering future queries.


\section*{Acknowledgments}
We wish to thank our anonymous reviewers for their constructive
suggestions which greatly improved this paper. This work was supported by the
National Science Foundation, NSF \#0423845 and NSF \#0438866, with
additional support from the ITIC KDD program.

\appendix
\section*{Appendix A}
\section*{Synthetic Data Generator}
\label{app:synth}

We have designed a synthetic data generator that allows us to
control different structural and attribute-based characteristics of
the data\shortcite{bhattacharya:tkdd07}. Here we present an overview of
the generation algorithm.

The generation process has two stages. In the first stage, we create
the collaboration graph among the underlying entities and the entity
attributes. In the second, we generate observed co-occurrence
relations from this collaboration graph. A high level description of
the generative process in shown in \figref{fig:synth-algo}.  Next,
we describe the two stages of the generation process in greater
detail.

The graph creation stage, in turn, has two sub-stages. First, we
create the domain entities and their attributes and then add
relationships between them. For creating entities, we control the
number of entities and the ambiguity of their attributes. We create
$N$ entities and their attributes one after another. For simplicity
and without losing generality, each entity $e$ has a single floating
point attribute $e.x$, instead of a character string. A parameter
$p_a$ controls the ambiguity of the entity attributes; with
probability $p_a$ the attribute of a new entity is chosen from
values that are already in use by existing entities. Then $M$ binary
relationships are added between the created entities. As with the
attributes, there is a parameter controlling the ambiguity of the
relationships, as defined in \secref{sec:rc-analysis}. For each
binary relationship ($e_i,e_j$), first $e_i$ is chosen randomly and
then $e_j$ is sampled so that $(e_i,e_j)$ is an ambiguous
relationship with probability $p^R_a$.

\begin{figure}[t]
\begin{minipage}[c]{5in}
{\small\sf
\begin{tabbing}
aaaaaaaaaaaa \= aaaaaa \= aaaaaaaa \= aaaaaa \=> \kill
 \> Creation Stage \\
1. \> Repeat N times \\
2. \> \> Create random attribute $x$ with ambiguity $p_a$ \\
3. \> \> Create entity $e$ with attribute $x$ \\
4. \> Repeat M times \\
5. \> \> Choose entity $e_i$ randomly \\
6. \> \> Choose entity $e_j$ with prob $p^R_a$ of an ambiguous relationship $(e_i,e_j)$  \\
7. \> \> Set $e_i = Nbr(e_j)$ and $e_j = Nbr(e_i)$ \\
\\
 \> Generation Stage \\
8. \> Repeat R times \\
9. \> \> Randomly choose entity $e$ \\
10. \> \> Generate reference $r$ using ${\mathcal N}(e.x, 1)$ \\
11.\> \> Initialize hyper-edge $h=\langle r\rangle$ \\
12.\> \> Repeat with probability $p_c$ \\
13.\> \> \> Randomly choose $e_j$ from $Nbr(e)$ without replacement \\
14.\> \> \> Generate reference $r_j$ using ${\mathcal N}(e_j.x, 1)$ \\
15.\> \> \> Add $r_j$ hyper-edge $h$ \\
16.\> \> Output hyper-edge $h$ \\
\end{tabbing}
\normalsize }
\end{minipage}
\caption[The synthetic data generation algorithm] {High-level
description of synthetic data generation algorithm}
\label{fig:synth-algo}
\end{figure}

Before describing the process of generating co-occurrence
relationships from the graph, let us consider in a little more
detail the issue of attribute ambiguity. What finally needs to be
controlled is the ambiguity of the reference attributes. While these
depend on the entity attributes, they are not completely determined
by entities. Taking the example of names, two people who have names
`John Michael Smyth' and `James Daniel Smith' can still be ambiguous
in terms of their observed names in the data depending on the
generation process of observed names. In other words, attribute
ambiguity of the references depends both on the separation between
entity attributes and the dispersion created by the generation
process. We make the assumption that for an entity $e$ with
attribute $e.x$, its references are generated from a Gaussian
distribution with mean $x$ and variance 1.0. So, with very high
probability, any reference attribute generated from $e.x$ will be in
the range $[e.x-3,e.x+3]$. So this range in the attribute domain is
considered to be `occupied' by entity $e$. Any entity has an
ambiguous attribute if its occupied range intersects with that of
another entity.

Now we come to the generation of co-occurrence relationships from
the entity collaboration graph. In this stage, $R$ co-occurrence
relationships or hyper-edges are generated, each with its own
references. For each hyper-edge $\langle
r_i,r_{i1},\ldots,r_{ik}\rangle$, two aspects need to be controlled
--- how many references and which references should be included in
this hyper-edge. This is done as follows. First, we sample an entity
$e_i$ which serves the initiator entity for this hyper-edge. Then
other entities $e_{ij}$ for this hyper-edge are repeatedly sampled
(without replacement) from the neighbors of the initiator entity
$e_i$. The size of the hyper-edge is determined using a parameter
$p_c$. The sampling step for a hyper-edge is terminated with
probability $p_c$ after each selection $e_{ij}$. The process is also
terminated when the neighbors of the initiator entity are exhausted.
Finally, references $r_{ij}$ need to be generated from each of the
selected entities $e_{ij}$. This is done for each entity $e$ by
sampling from its Gaussian distribution ${\mathcal N}(e.x,1)$.

\vskip 0.2in

\end{document}